# Structural phase transition and bandgap control through mechanical deformation in layered semiconductors 1T-$ZrX_2$ (X = S, Se)

*Edoardo Martino[1]*, David Santos-Cottin[2], Florian Le Mardelé[2], Konstantin Semeniuk[1], Michele Pizzochero[1], Kristiāns Čerņevičs[1], Benoît Baptiste[3], Ludovic Delbes[3], Stefan Klotz[3], Francesco Capitani[4], Helmuth Berger[1], Oleg V. Yazyev[1] and Ana Akrap[2]**

[1] Institute of Physics, Ecole Polytechnique Fédérale de Lausanne (EPFL), CH-1015 Lausanne, Switzerland
[2] Department of Physics, University of Fribourg, CH-1700 Fribourg, Switzerland
[3] Institut de Minéralogie, de Physique des Matériaux et de Cosmochimie, Sorbonne University, CNRS UMR 7590, IMPMC, F-75005, Paris, France
[4] Synchrotron-SOLEIL, Saint-Aubin, BP48, F-91192 Gif-sur-Yvette Cedex, France

E-mail: edoardo.martino@epfl.ch, ana.akrap@unifr.ch

Applying elastic deformation can tune a material's physical properties locally and reversibly. Spatially modulated lattice deformation can create a bandgap gradient, favouring photo-generated charge separation and collection in optoelectronic devices. These advantages are hindered by the maximum elastic strain that a material can withstand before breaking. Nanomaterials derived by exfoliating transition metal dichalcogenides (TMDs) are an ideal playground for elastic deformation, as they can sustain large elastic strains, up to a few percent. However, exfoliable TMDs with highly strain-tunable properties have proven challenging for researchers to identify.

We investigated 1T-$ZrS_2$ and 1T-$ZrSe_2$, exfoliable semiconductors with large bandgaps. Under compressive deformation, both TMDs dramatically change their physical properties. 1T-$ZrSe_2$ undergoes a reversible transformation into an exotic three-dimensional lattice, with a semiconductor-to-metal transition. In $ZrS_2$, the irreversible transformation between two different layered structures is accompanied by a sudden 14% bandgap reduction. These results establish that Zr-based TMDs are an optimal strain-tunable platform for spatially textured bandgaps, with a strong potential for novel optoelectronic devices and light harvesting.

When developing semiconductor-based devices for photovoltaic, photocatalytic or photosensing applications, light harvesting efficiencies can be optimised by fine-tuning the size, or even the spatial variation of the electronic bandgap. This can be achieved by mechanical actions, such as strain or hydrostatic pressure. Application of high pressure is an experimentally accessible path for exploring how the properties of a material change upon varying the interatomic distances [1].

Established techniques, such as X-ray diffraction (XRD), can then be used in combination to thoroughly quantify the response of crystalline systems to various mechanical actions, and compute the corresponding compressibility and elastic moduli.

Once a material's mechanical and electronical response to applied forces is known, one can design a device in which, for example, a specific strain induced in the photosensitive component attunes it to a chosen wavelength. A variety of tools and methods for such strain-tuning have been implemented and reported, such as local compression by AFM cantilevers [2], adhesion to an elastomeric substrate [3,4], patterned substrates [5,6] and the use of differential thermal expansion between the active material and the substrate [7].

Layered van der Waals materials are especially good candidates for strain-engineered optoelectronic devices, thanks to their proven extremely wide range of elastic strain which they can support (above 5%), when reduced to nanometre thickness by exfoliation [8]. The strain tuning of optoelectronic properties of layered materials, such as transition metal dichalcogenides (TMDs), has already been proposed and successfully demonstrated for the semiconducting compounds containing group 6 metals: 2H-$MoS_2$ and 2H-$WSe_2$ [9-11]. Furthermore, tuning the bandgap in TMDs by mechanical action has been used to enhance the performance of semiconducting devices for light harvesting applications [12]. The prototypical 2H-$MoS_2$ and 2H-$WSe_2$ are, however, not the only semiconducting layered TMDs. These also include systems with group 4 metals, as Zr and Hf. The corresponding bandgaps range from 1.2 eV to 2 eV, pertinent for optoelectronic and energy conversion applications. These compounds have so far been neglected in comparison to the Mo- and W-based semiconductors, despite their rather promising properties which encompass a high electronic mobility [13] and a tendency to spontaneously form native oxides with high dielectric constant [14]. The salient differences between group 4 and group 6 TMDs are in the electronic energy bands' orbital character [15]. Group 4 TMDs have two fewer valence electrons on the metal atom, and a different metal-chalcogen coordination. Group 6 compounds—for example $MoS_2$—crystallize in the 2H polytype, where the metal is in a trigonal prismatic coordination with six chalcogen atoms. Both conduction and valence bands have a metallic, $d$ orbital character, giving the chalcogen a minor role in the electronic structure. In contrast, Zr- and Hf-based TMDs assume the 1T-polytype structure with a six-fold octahedral metal-chalcogen coordination. Their conduction bands originate from the transition metal $d$ orbitals, while the chalcogen $p$ orbitals are responsible for the valence bands.

Realising a structural phase transition driven by applying a mechanical action is a long-standing desire of the community working on layered semiconductors, such as TMDs. A mechanically induced lattice reconstruction in TMDs can open a venue for strain-controlled semiconductor-to-metal conversion [16]. This phenomenon can be utilised in tunable optoelectronic elements, or in clean metal-semiconductor lateral homojunctions [17,18], operated by a stretchable substrate. However, it is not possible to mechanically induce phase switching in 2H-$MoS_2$ and 2H-$WSe_2$. In fact, structural changes in these compounds cannot be triggered by strain or pressure alone [19-22] and require additional high temperatures, laser irradiation or intercalation [23,24]. In the present study, we investigated 1T-$ZrS_2$ and 1T-$ZrSe_2$, motivated by the 1T polytypes being considerably more prone to structural instabilities than the 2H polytypes. In light of the different chalcogen size for the investigated materials, Se being larger than S, their response under applied pressure may be expected to vary. For both mentioned compounds, we observed transitions between two distinct crystalline structures, characterized by different electronic properties. Such lattice reconstructions commonly occur in metallic TMDs, and are responsible for the unusual transport properties of 1T-$TaS_2$ [25], and topological phase transitions in 1T-$MoTe_2$ [26]. The transition temperatures at which these transformations manifest can be tuned in a broad temperature range by applied pressure [27-29]. Indeed, by means of *ab* initio calculations, Zhai et al. [30], predicted a structural phase transition to occur in 1T-$ZrS_2$ at high pressure. To test this experimentally in 1T-$ZrS_2$ and 1T-$ZrSe_2$, we measured their optical and structural properties under quasi-hydrostatic pressure that we applied using a diamond anvil cell.

**Results**

Figure 1 summarizes a set of structural, infrared and Raman properties for 1T-$ZrS_2$ and 1T-$ZrSe_2$ at ambient pressure. The data were recorded on the same samples as those later used for the high-pressure experiments, and are consistent with the published results [32]. Optical transmission measurements (Figure 1e) show transparency in a wide energy range, and the onset of optical absorption at the indirect bandgap energies, which are 1.7 eV for $ZrS_2$ and 1.2 eV for $ZrSe_2$. Transparency in the infrared range is accompanied by a remarkably high room-temperature dc conductivity of 3.13 S/cm (0.31 Ωcm) for $ZrS_2$, and 23.81 S/cm (0.04 Ωcm) for $ZrSe_2$. The high intrinsic doping level of these materials [31, 32], causing their high dc conductivity, is also evidenced by the intense light absorbance at low energies (< 0.5 eV). This is most likely due to in-gap or shallow donor states.

Single crystals of Zr-based TMDs were loaded into a diamond anvil cell for optical measurements under high-pressure conditions. A finely ground powder was used for X-ray diffraction experiments. The applied pressure was monitored using ruby fluorescence [33]. The compressive strain imposed onto the materials at every specific pressure was calculated from our data, and can be found in the SI. The pressure dependence of the Raman and powder XRD patterns are shown in Figure 2. For both materials, the formations of different crystalline structures are evident from the appearance of new peaks in the Raman and XRD spectra. In $ZrS_2$, the new crystalline phase starts to appear at 3 GPa, which corresponds to 1% in-plane and 4% out-of-plane compression (Figure SI. 3). With increasing pressure, a larger fraction of the sample is converted into this new phase, with the low- and high-pressure phases coexisting over a broad range of pressures. The transformation into the newly created phase, here called HP-$ZrS_2$, is irreversible upon full decompression to 1 atm. $ZrSe_2$ remains structurally stable up to 8 GPa, equivalent to 3% in-plane and 7% out-of-plane compression (Figure SI. 3). Upon further increase of pressure, a structural transformation takes place, with a more rapid conversion of the material compared to the S-based compound. The high-pressure phase (HP-$ZrSe_2$) persists down to 2 GPa when pressure is released. Below 2 GPa, the initial structure is recovered, making the transformation reversible, but with a wide hysteresis. The progression and reversibility of the phase transitions as a function of pressure are schematically depicted in Figure 2 g,h. The different behaviour between the two isostructural materials—different transformation onset pressures as well as the presence or absence of reversibility—suggests a possible difference in the crystalline structure of the high-pressure phases. It is not yet clear what makes this structural transformation favourable only for these specific types of TMDs, but we believe that the different metal-chalcogen coordination, when compared to 2H-polytypes, and the fact that there are two valence electrons less [15] on the transition metal, must play a fundamental role.

Using our high-pressure XRD and Raman data, and aided by the work of Zhai et al. [30], we succeeded in refining the crystalline structures of HP-$ZrS_2$ and HP-$ZrSe_2$. For $ZrS_2$, mechanical deformation transforms the known layered material into a new layered structure (Fig. 3 a), previously never observed in a TMD. The octahedra are transformed into distorted trigonal prisms. The structure is different from that of a 2H phase, as the triangular bases of the prisms are now parallel to the *bc*-plane, and their orientation alternates in the *a*-axis direction. Even more surprising and unusual is the structure assumed by $ZrSe_2$. This initially layered material becomes three-dimensional, a transformation that is reminiscent of the graphite to diamond conversion. The resultant crystalline structure—orthorhombic with the space group *Immm*—is unusual, and can be observed at ambient conditions only in exotic materials, such as the rare-earth dichalcogenide $ErSe_2$ [34] and the unconventional heavy-fermion superconductor $UTe_2$ [35]. The Zr atoms are coordinated by 8 chalcogens, a configuration that can also be found at ambient pressure in the quasi-one-dimensional crystals $ZrS_3$ and $ZrSe_3$ [36].

Rietveld refinement of the powder X-ray diffraction patterns was done for every pressure point (Figures SI. 1 and 2). This was used to determine the possible crystal structure for the high-pressure phases, along with the compressibility and volume fraction of the 1T phase as a function of pressure.

Having identified the crystalline structures, the optical and electronic properties of those newly created materials are of immediate interest. By measuring optical transmittance as a function of pressure, we investigated the presence and magnitude of the optical bandgap (further details can be found in the SI). For $ZrS_2$, at low pressures (0 to 2 GPa) the bandgap of the 1T phase is reduced, like it is commonly seen in other TMDs [19,20]. Above 3 GPa, as the new phase appears, it is possible to identify the contribution to absorption from a second bandgap, which is approximately 200 meV smaller than that of the native 1T phase. As pressure further increases, the bandgap decreases, offering additional control over the optical properties. At the same time, the volume ratio between the two phases changes, as seen from the XRD data (SI), further modifying the optical spectrum. To confirm our estimate of the optical bandgap, we performed first-principle calculations on the $ZrS_2$ HP phase at 6 GPa. The computed band structure has an optical bandgap of 1 eV, in perfect agreement with the value extracted from the experimental data (Figure SI 8). The origin of a small suppression of optical transmittance in the 0.5 – 1 eV energy range remains unclear.

$ZrSe_2$ changes its properties more dramatically. We observed a continuous reduction of the bandgap of the 1T phase up to 5 GPa,. As the new crystalline phase appears, the optical transparency is suppressed for all frequencies. This means that HP-$ZrSe_2$ is metallic, or possibly a narrow-gap semiconductor with a bandgap of less than 0.1 eV. The evolution of the optical bandgaps extracted from the optical transmittance data is summarized in Figure 3 c,f.

In order to further address the dramatic pressure-induced semiconductor-to-metal transition observed in $ZrSe_2$, the band structure for the proposed high-pressure phase was computed by means of semilocal Perdew-Burke-Ernzerhof functional. Specifically, we determined the electronic band structures of both the low-pressure (0 GPa) and the high-pressure (9.5 GPa) phases by means of density functional theory. The low-pressure phase is found to be a semiconductor, with an indirect bandgap of 0.9 eV (see Figure 4a). This value is in agreement with our optical measurements shown in Figure 3e and 3f, and perfectly matches the results of recent ARPES experiments [31]. We found that the valence band maximum is located at the centre of the Brillouin zone, while the conduction band minimum occurs at the high-symmetry L point. As far as the high-pressure phase of $ZrSe_2$ is concerned, the present optical measurements cannot distinguish whether this system is a metal, or a semiconductor with a bandgap below 0.1 eV. Our calculations reported in Figure 4b clearly indicate a metallic character, with a number of bands crossing the Fermi level. Overall, our calculations paint a picture in accordance with the experimental observations.

To conclude, we have experimentally demonstrated room-temperature pressure-driven structural phase transitions in the semiconducting layered transition metal dichalcogenides 1T-$ZrS_2$ and 1T-$ZrSe_2$. The two materials are isostructural and isoelectronic, but undergo completely different lattice reconstructions, and their physical properties evolve in distinct manners. Under applied pressure of 3 GPa, $ZrS_2$ is transformed irreversibly into a new layered structure, with a consequent reduction of the bandgap by 200 meV. As for $ZrSe_2$, once pressure is above 8 GPa, the layers merge into a 3D structure. This collapse results in a semiconductor-to-metal transition, as confirmed by *ab initio* calculations. In addition, the bandgap of the starting 1T phase reduces under compression at the rate of 50 meV/GPa, giving a further tunability over the materials' optical properties. Our demonstration of structural transformations induced by mechanical actions, with the associated dramatic changes in their electronic properties, grants the long-awaited additional functionality in TMDs, enabling us to control structure and properties on-demand. New opportunities in the field of strain-designed devices—charge funnelling and clean metal-semiconductor homojunctions in optoelectronic components—are now within reach. Further research needs to be done in the direction of understanding the dynamics of the lattice reconstruction, in particular for $ZrSe_2$ because of its close reminiscence to the graphite to diamond transformation.

## Methods

**Crystals synthesis:** Single crystals of 1T-$ZrS_2$ and 1T-$ZrSe_2$ were synthesized by chemical vapour transport method using $I_2$ as transport agent. Similar conditions as those presented in [32] have been used to growth large single crystals.

**High-pressure Raman and infrared spectroscopy:** Raman and infrared measurements were performed on the infrared beamline SMIS at the synchrotron SOLEIL. We used a DXR2 Thermo Fisher Raman microspectrometer equipped with a 633nm He-Ne laser as light source, a Peltier cooled CCD detector, a 600 grooves/mm grating, resulting in a spectral resolution of about 4 $cm^{-1}$. Infrared measurements were performed on a custom-built horizontal microscope for diamond anvil cells (DACs), equipped with custom Cassegrain objectives, a MCT and a Si diode detectors, for the Mid-IR and the Near-IR respectively [37]. The horizontal microscope was coupled to Thermo Fisher iS50 interformeter with Quartz (NIR) / KBr (MIR) beamsplitters and synchrotron radiation as IR source.

Square samples of about 100 × 100 $\mu m^2$ were loaded into a membrane DAC with 400 µm culets. Stainless-steel gaskets were pre-indented to a thickness of about 50 µm and a hole of diameter 150 um was drilled by electro-erosion. NaCl was used as pressure transmitting medium [38]. Pressure was measured in situ by the standard ruby fluorescence technique. Data was collected every 1 GPa during both compression and decompression, with the last data point collected after the cell was  open to release any residual pressure.

**X-ray diffraction:** The high-pressure X-ray diffraction experiments were performed at the X-ray diffraction platform of IMPMC at 0.85GPa and 293K, using a membrane diamond anvil cell (DAC) with a culet of 500 μm in diameter. A 250 μm-diameter hole was drilled in a stainless-steel gasket with a thickness of 200 μm pre-indented to 80 μm. The finely ground powders of $ZrSe_2$ and $ZrS_2$ were loaded into the gasket hole together with a ruby ball as pressure sensor [39] (R1 ruby fluorescence method). A 4:1 methanol-ethanol mixture was used as pressure transmitting fluid since it is known to provide hydrostatic conditions to approximately 10 GPa [40]. Before and after each measurement, pressure was determined by recording the ruby fluorescence spectra using a spectrometer with a 532 nm excitation laser. The calculated error for the pressure values is 0.1 GPa and the sensitivity was 0.05GPa.

The DAC was then mounted on a Rigaku MM007HF diffractometer equipped with a Mo rotating anode ($\lambda_{K\alpha1}$ = 0.709319 Å, $\lambda_{K\alpha2}$ = 0.713609 Å), VariMax focusing optics and a RAXIS4++ image plate detector. X-ray data were collected at 20 °C. A $LaB_6$ standard sample was measured in the same experimental conditions to calibrate the FIT2D program, the image processing software used to integrate the intensities around the Debye−Scherrer rings and to get the 1D patterns.

**Density Functional Theory Calculations***:* Our first-principles calculations were performed at the density functional theory level as implemented in VASP [42,43]. We adopted the hybrid HSE06 [41] functional for describing the semiconducting systems, and the semilocal PBE functional for the metallic one [44]. Electron-core interactions were described through the projector augmented wave method, while Kohn-Sham wavefunctions were expanded in a plane wave basis set with a cutoff on kinetic energy of 400 eV. The integration over the Brillouin zone was carried out using a 6×6×4 (6×4×2) k-mesh for the low-pressure (high-pressure) crystalline phase of $ZrSe_2$. We relied on the experimental crystal structures determined in our X-ray diffraction experiments.

**Acknowledgements**

The authors acknowledge illuminating discussions with Laszlo Forró, Nathaniel Miller, Anna Celeste, Ferenc Borodincs, Quansheng Wu, Zoran Rukelj and Alexey Kuzmenko. We would like to thank Philippe Rosier and Nicolas Dumesnil (from IMPMC) for the production of mechanical parts which were useful for mounting the diamond anvil cell on the diffractometer and high-pressure X-ray diffraction experiments.
E.M. and K.S. acknowledges funding from the Swiss National Science Foundation through its SINERGIA network MPBH and grant No. 200021_175836.
A.A. acknowledges funding from the Swiss National Science Foundation trough project PP00P2_170544.

M.P., K.C. and O.V.Y. are financially supported by the Swiss National Science Foundation through the grant 172543. First-principles calculations were performed at the Swiss National Supercomputing Centre (CSCS) under the project s832.
Experiment at Synchrotron Soleil was supported by the grant proposal ID 20190927.

## Author contribution

E.M. conceived the project and designed the experiments. H.B. grew the samples. E.M, D.S.C, F.L.M., K.S. and F.C performed the high-pressure Raman and infrared spectroscopy experiments. S.K., L.D. and B.P. performed the high-pressure X-ray diffraction experiments. D.S.C., E.M. S.K., L.D. and B.P. analysed and refined the X-ray diffraction patterns.
K.C., M.P, and O.V.Y. performed DFT calculations of band structure. A.A. supervised the project. E.M. wrote the paper, and all authors commented on the manuscript.

## Competing interests

The authors declare no conflict of interest.

## Data availability

The data that support the findings of this study are available from the authors (E.M. and A.A.) upon reasonable request.

## References

[1] Willardson, R., Weber, E., Paul, W. & Suski, T. High pressure semiconductor physics I. Academic Press, 1998.
[2] Manzeli, S., Allain, A., Ghadimi, A. & Kis, A. Piezoresistivity and strain-induced band gap tuning in atomically thin MoS2. Nano Lett. **2015**, 15, 5330–5335.
[3] Castellanos-Gomez, A. et al. Local strain engineering in atomically thin MoS2. Nano Lett. **2013**, 13, 5361–5366.
[4] Zhang, Q. et al. Strain relaxation of monolayer WS2 on plastic substrate. Adv. Funct. Mater. **2016**, 26, 8707–8714.
[5] Levy, N. *et al.* Strain-induced pseudo–magnetic fields greater than 300 Tesla in graphene nanobubbles. *Science.* **2010**, 329, 544–547.
[6] Li, H. et al. Optoelectronic crystal of artificial atoms in strain-textured molybdenum disulphide. Nat. Commun. **2015**, 6, 7381.
[7] Frisenda, R. et al. Biaxial strain tuning of the optical properties of single-layer transition metal dichalcogenides. npj 2D Mater. Appl. **2017**, 1, 10.
[8] Deng, S., Sumant, A. V. & Berry, V. Strain engineering in two-dimensional nanomaterials beyond graphene. Nano Today. **2018**, 22, 14–35.
[9] Feng, J., Qian, X., Huang, C. W. & Li, J. Strain-engineered artificial atom as a broad-spectrum solar energy funnel. Nat. Photonics. **2012**, 6, 866–872.
[10] Xie, S. et al. Coherent, atomically thin transition-metal dichalcogenide superlattices with engineered strain. Science. **2018**, 359, 1131–1136.
[11] Conley, H. J. *et al.* Bandgap engineering of strained monolayer and bilayer MoS2. *Nano Lett.* **2013**, 13, 3626–3630.
[12] De Sanctis, A., Amit, I., Hepplestone, S. P., Craciun, M. F. & Russo, S. Strain-engineered inverse charge-funnelling in layered semiconductors. Nat. Commun. **2018**, 9, 1652.
[13] Zhang, W., Huang, Z., Zhang, W. & Li, Y. Two-dimensional semiconductors with possible high room temperature mobility. *Nano Res.* **2014**, **7**, 1731–1737.

[14] Lai, S. *et al.* HfO 2/HfS 2 hybrid heterostructure fabricated via controllable chemical conversion of two-dimensional HfS 2. *Nanoscale.* **2018**, 10, 18758–18766.
[15] Kolobov, A. V & Tominaga, J. Two-Dimensional Transition-Metal Dichalcogenides. 239, *Springer*, 2016.
[16] Duerloo, K.-A. N., Li, Y. & Reed, E. J. Structural phase transitions in two-dimensional Mo-and W-dichalcogenide monolayers. *Nat. Commun.* **2014**, **5**, 4214.
[17] Eda, G. *et al.* Coherent atomic and electronic heterostructures of single-layer MoS 2. *ACS Nano* **2012**, 6, 7311–7317.
[18] Kappera, R. *et al.* Phase-engineered low-resistance contacts for ultrathin MoS2 transistors. *Nat. Mater.* **2014**, 13, 1128–1134.
[19] Nayak, A. P. et al. Pressure-induced semiconducting to metallic transition in multilayered molybdenum disulphide. *Nat. Commun*. **2014**, 5, 3731.
[20] Zhao, Z. et al. Pressure induced metallization with absence of structural transition in layered molybdenum diselenide. *Nat. Commun*. **2015**, 6, 7312.
[21] Wang, X. et al. Pressure-induced iso-structural phase transition and metallization in WSe2. *Sci. Rep*. **2017**, 7, 46694.
[22] Caramazza, S. *et al.* Effect of pressure on optical properties of the transition metal dichalcogenide MoSe2. in *Journal of Physics: Conference Series*, **2017**, 950.
[23] Guo, Y. *et al.* Probing the dynamics of the metallic-to-semiconducting structural phase transformation in MoS2 crystals. *Nano Lett.* **2015**, 15, 5081–5088.
[24] Wang, L., Xu, Z., Wang, W. & Bai, X. Atomic mechanism of dynamic electrochemical lithiation processes of MoS2 nanosheets. *J. Am. Chem. Soc.* **2014**, 136, 6693–6697.
[25] Martino, Edoardo, et al. "Preferential out-of-plane conduction and quasi-one-dimensional electronic states in layered van der Waals material 1T-TaS2." *npj 2D Materials and Applications*, **2020**, 4.1: 1-9.
[26] Wang, Zhijun, et al. "MoTe 2: a type-II Weyl topological metal." *Physical review letters*. **2016**, 117.5, 056805.
[27] Sipos, B. *et al.* From Mott state to superconductivity in 1T-TaS 2. *Nat. Mater.* **7**, 960 (2008).
[28] Berger, A. N. *et al.* Temperature-driven topological transition in 1T'-MoTe 2. *npj Quantum Mater.* **2018**, 3, 2.
[29] Heikes, C. *et al.* Mechanical control of crystal symmetry and superconductivity in Weyl semimetal MoTe 2. *Phys. Rev. Mater.* **2018**, 2, 74202.
[30] Zhai, H. *et al.* Pressure-induced phase transition, metallization and superconductivity in ZrS 2. *Phys. Chem. Chem. Phys.* **2018**, 20, 23656–23663.
[31] Ghafari, A., Moustafa, M., Di Santo, G., Petaccia, L. & Janowitz, C. Opposite dispersion bands at the Fermi level in ZrSe2. *Appl. Phys. Lett.* **2018**, 112, 182105.
[32] Moustafa, M., Zandt, T., Janowitz, C. & Manzke, R. Growth and band gap determination of the ZrSx Se2-x single crystal series. *Phys. Rev. B - Condens. Matter Mater. Phys.* **2009**, 80, 035206-1–6.
[33] Mao, H. K., Bell, P. M., Shaner, J. W. & Steinberg, D. J. Specific volume measurements of Cu, Mo, Pd, and Ag and calibration of the ruby R1 fluorescence pressure gauge from 0.06 to 1 Mbar. *J. Appl. Phys.* ***1978***, 49, 3276–3283.
[34] Haase, D. J., Steinfink, H. & Weiss, E. J. The Phase Equilibria and Crystal Chemistry of the Rare Earth Group VI Systems. I. Erbium-Selenium. *Inorg. Chem.* **1965**, 4, 538–540.
[35] Aoki, D. *et al.* Unconventional Superconductivity in Heavy Fermion UTe2. *J. Phys. Soc. Japan*. **2019**, 88, 43702.
[36] Furuseth, S., Brattas, L. & Kjekshus, A. Crystal Structures of TiS 3, ZrS 3, ZrSe 3, ZrTe 3, HfS 3 and HfSe 3. *Acta Chem. Scand.* **1975**, 29, 623.
[37] Livache, C. *et al.* Effect of Pressure on Interband and Intraband Transition of Mercury Chalcogenide Quantum Dots. *J. Phys. Chem. C.* **2019**, 123, 13122–13130.

[38] Celeste, A., Borondics, F. & Capitani, F. Hydrostaticity of pressure-transmitting media for high pressure infrared spectroscopy. *High Press. Res.* **2019**, 39, 608–618.
[39] Chervin, J. C., B. Canny, and M. Mancinelli. "Ruby-spheres as pressure gauge for optically transparent high pressure cells." *Inter. J. of High Pressure Research.* **2001**, 21.6, 305-314.
[40] Klotz, S., et al. "Hydrostatic limits of 11 pressure transmitting media." *J. of Physics D: Applied Physics.* **2009**, 42.7, 075413.
[41] Heyd, J., Scuseria, G. E. & Ernzerhof, M. Hybrid functionals based on a screened Coulomb potential. *J. Chem. Phys.* **2003**, 118, 8207–8215.
[42] Joubert, D. From ultrasoft pseudopotentials to the projector augmented-wave method. *Phys. Rev. B - Condens. Matter Mater. Phys.* **1999**, 59, 1758–1775.
[43] Kresse, G. & Furthmüller, J. Efficient iterative schemes for ab initio total-energy calculations using a plane-wave basis set. *Phys. Rev. B - Condens. Matter Mater. Phys.* **1996**, 54, 11169–11186.
[44] Perdew, J. P., Burke, K. & Ernzerhof, M. Generalized gradient approximation made simple. *Phys. Rev. Lett.* **1996**, **77**, 3865–3868.

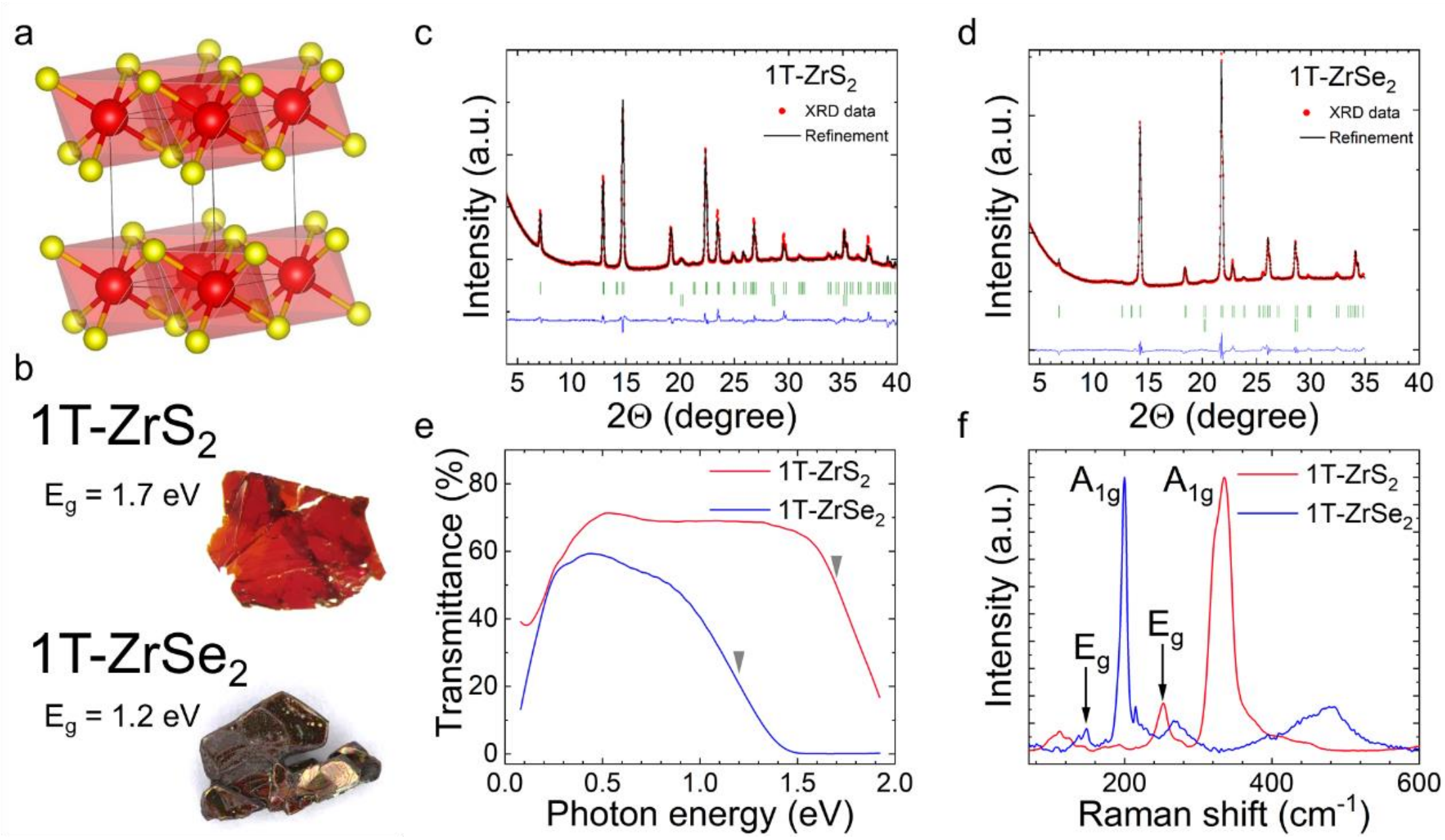

**Figure 1. Structural and optical properties of 1T-$ZrS_2$ and 1T-$ZrSe_2$. a**) Crystal structure of both $ZrS_2$ and $ZrSe_2$. Red and yellow spheres represent the metal and chalcogen atoms, respectively. The octahedra formed by the chalcogens around the central Zr atom are shown. **b**) Photographs of the single crystals used for the experiments. The bright red colour of $ZrS_2$ is the result of its optical bandgap in the visible range. **c,d**) Rietveld-refined powder X-ray diffraction (XRD) of 1T-$ZrS_2$ and 1T-$ZrSe_2$ measured at low pressure (< 1 GPa) inside the diamond anvil cell. Red dots are the experimental data, and the black lines are the simulated patterns for the known crystal structures. The diffraction pattern from the steel gasket is also included into the refinement. **e**) Optical transmittance through thin samples (6 μm and 2 μm for $ZrS_2$ and $ZrSe_2$, respectively), where the arrows indicate the positions of the indirect bandgaps at the onset of optical absorption. **f**) Raman spectra with the indexed $A_{1g}$ and $E_g$ phonon modes, shared by both materials due to their identical crystalline symmetry.

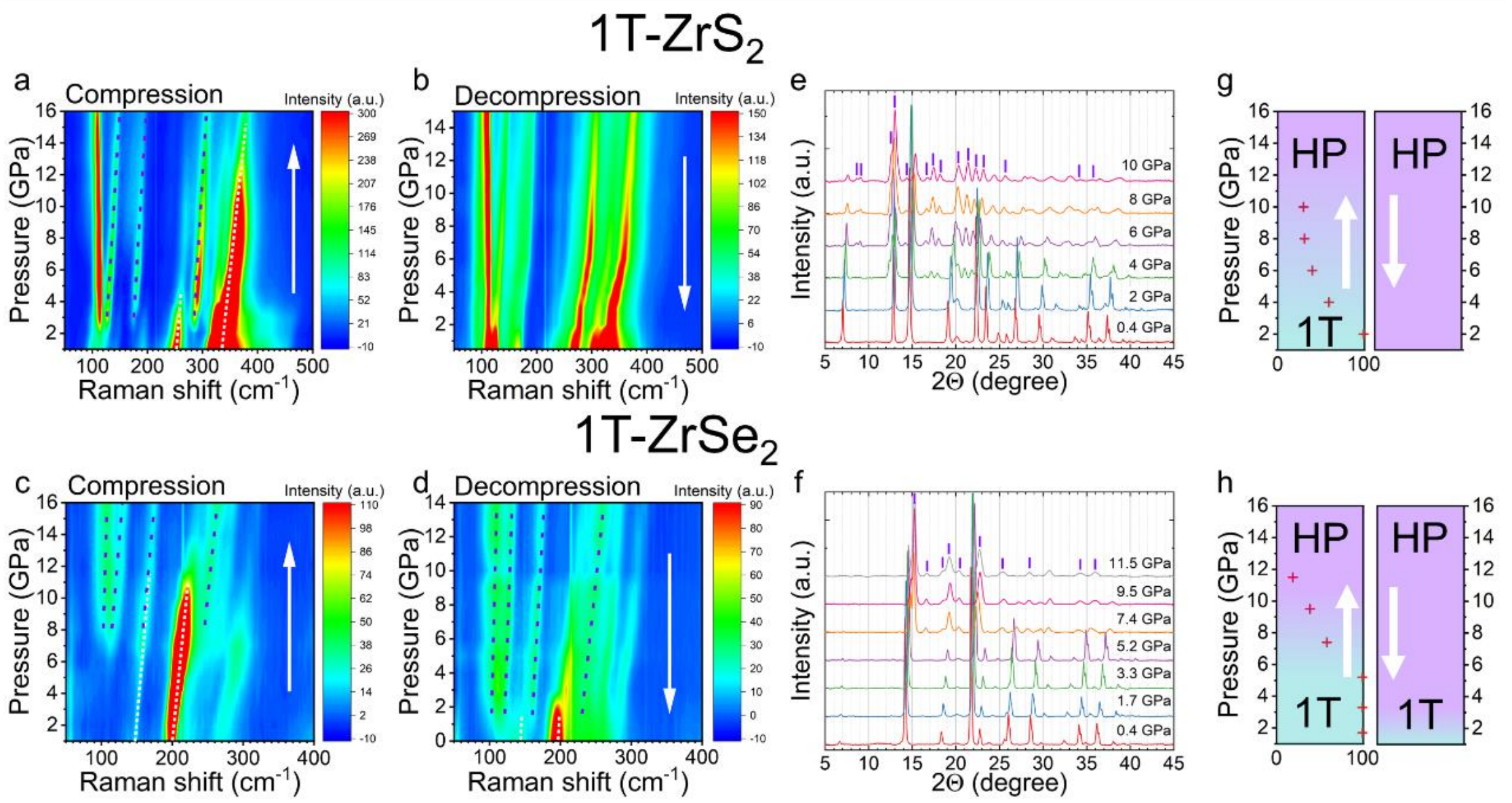

**Figure 2. Evidence of pressure-induced structural phase transitions a,b**) Raman spectra of $ZrS_2$ during compression (**a**) and decompression (**b**). The dotted white and purple lines are guides for the eyes in the low- and high-pressure phases, respectively. The compression curve shows the onset of transformation at 3 GPa. During decompression the high-pressure phase remains stable down to 1 atm. **c,d**) Pressure evolution of the Raman spectra of $ZrSe_2$. Data during compression (**c**) shows the appearance of a new crystalline phase at 8 GPa, and the initial phase is restored during decompression below 2 GPa (**d**). **e,f**) Powder X-ray diffraction at different pressures for $ZrS_2$ (**e**) and $ZrSe_2$ (**f**), the purple lines mark the diffraction peaks originating from the new crystalline phase. **g,h**) Schematic representation of the pressure driven transformation. The blue and purple regions represent the ambient-pressure and high-pressure phases, respectively. The red crosses show the volume fraction of the 1T phase as a function of pressure during compression.

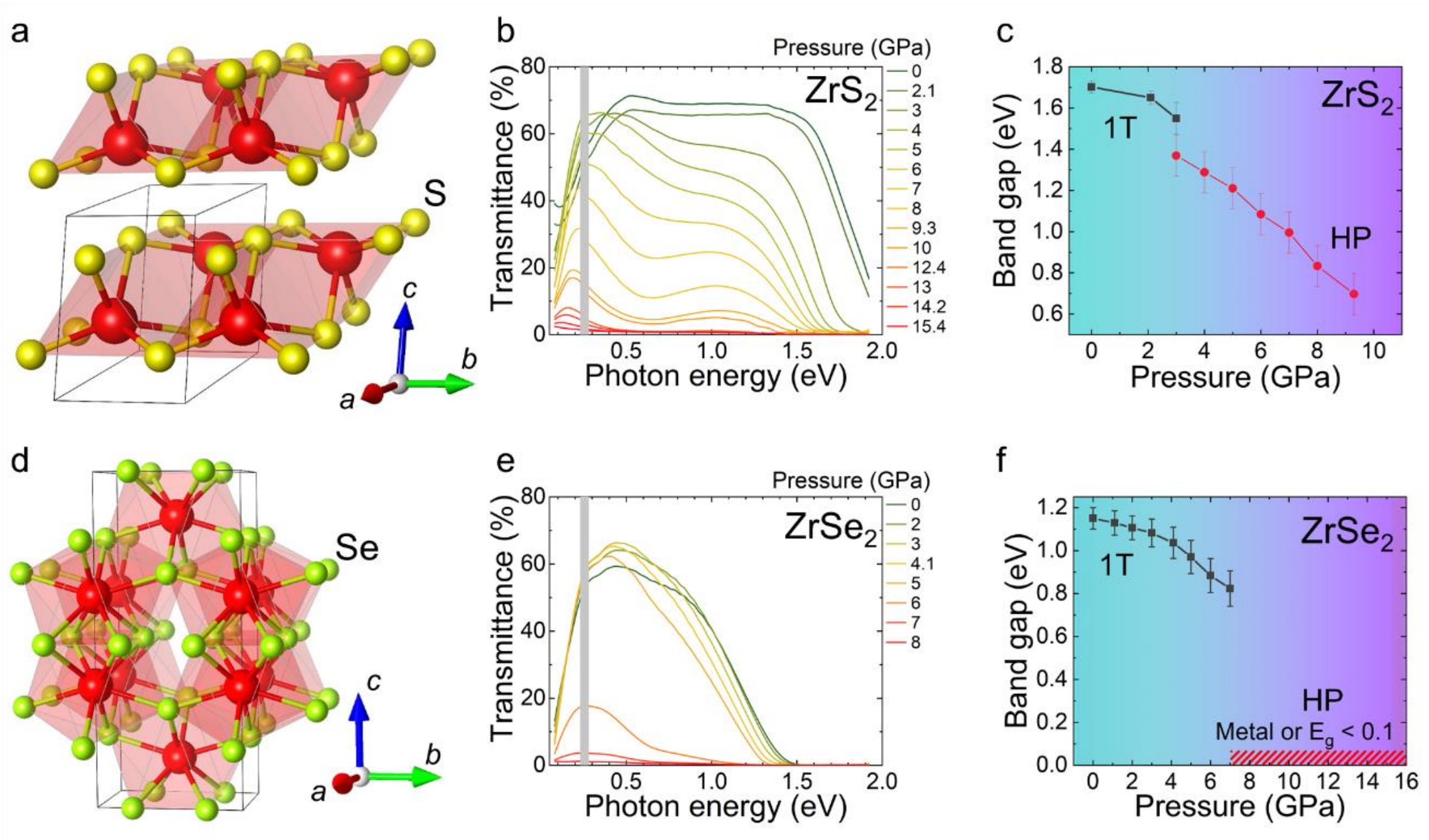

**Figure 3. High-pressure structures and switching of optical properties a**) Crystal structure of HP-$ZrS_2$. The layered structure is evident, while the original octahedral coordination is distorted into an unusual trigonal prismatic configuration. **b**) Optical transmittance as function of pressure for $ZrS_2$. **c**) $ZrS_2$ bandgap evolution under pressure **d**) Crystal structure of HP-$ZrSe_2$. The atomic lattice is three-dimensional, a rare configuration that exists at ambient pressure in $ErSe_2$ and $UTe_2$. **e**) Optical transmittance measured up to 8 GPa. As the new phase appears, optical transmittance becomes zero due to its metallic properties. **f**) $ZrSe_2$ bandgap evolution under pressure. The grey area in the panels **b** and **e** are the 0.23-0.28 eV region obscured by diamond absorption.

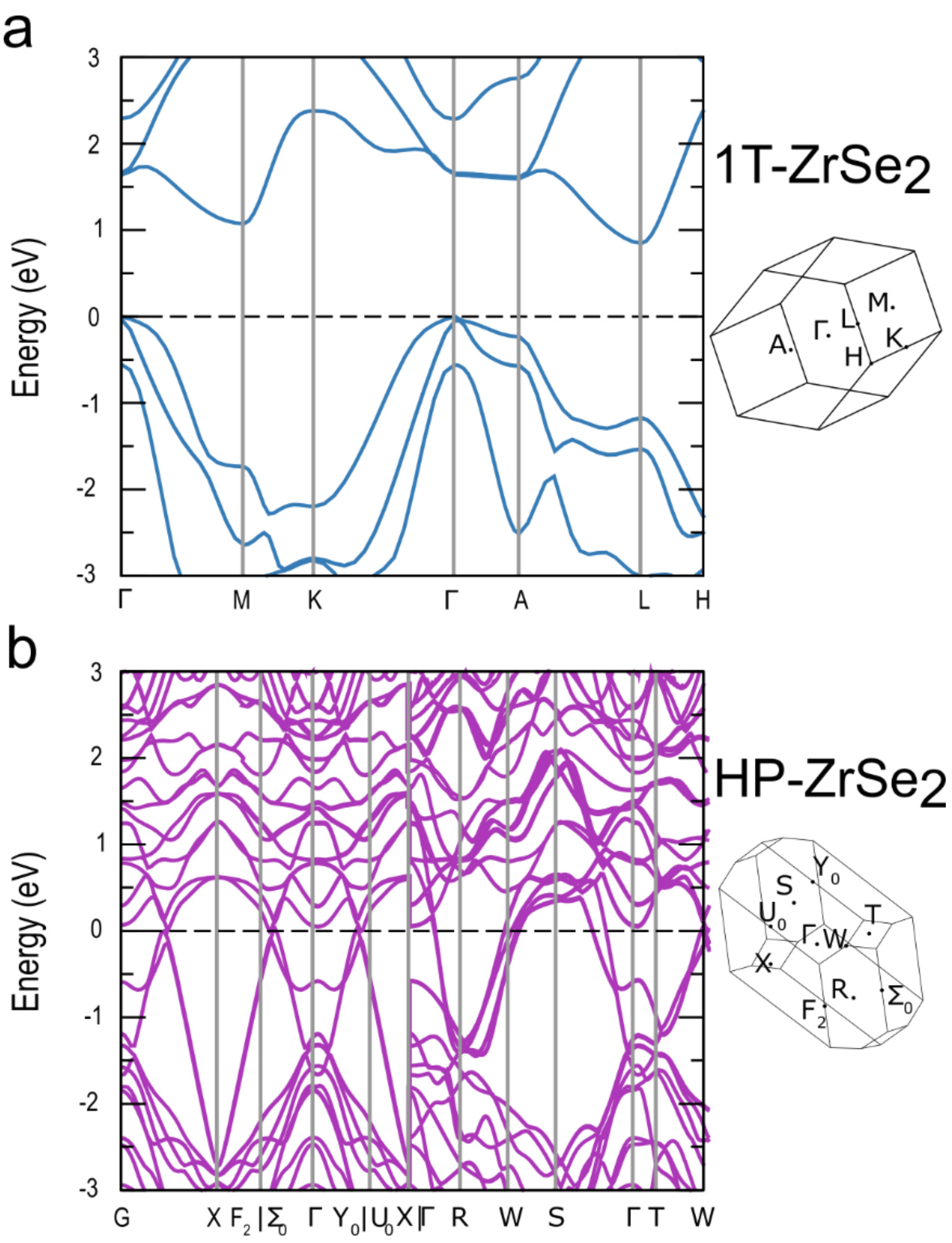

**Figure 4. First-principles calculations of the electronic structure of $ZrSe_2$.** Electronic band structure of the (**a**) low-pressure and (**b**) high-pressure crystalline phases of $ZrSe_2$, together with the high-symmetry points in the corresponding first Brillouin zones.

# Supporting Information

## Pressure-driven Structural Phase Transition and Band-gap Collapse in Layered Semiconductors 1T-$ZrX_2$ (X = S,Se)

*Edoardo Martino*, David Santos-Cottin, Florian Le Mardelé, Konstantin Semeniuk, Michele Pizzochero, Kristiāns Čerņevičs, Stefan Klotz, Ludovic Delbes, Benoît Baptiste, Francesco Capitani, Helmuth Berger, Oleg V. Yazyev and Ana Akrap**

### High-pressure Rietveld refinement of the XRD spectra

The crystal structure refinements were carried out using the Rietveld method as implemented in the Fullprof software [1].

Starting from the known cell parameters of $ZrS_2$ and $ZrSe_2$, the refinements converged rapidly for the known structures at low pressure, considering low amounts of iron (diffraction from the DAC gasket).

For the two structures, preferred orientations were evidenced and their relative parameters refined: (001) for 1T-$ZrS_2$, (-101) for HP-$ZrS_2$, (001) for 1T-$ZrSe_2$, and (010) for HP-$ZrSe_2$.

The peak widths were significantly larger than the instrument resolution (~0.15° 2θ). Starting from an instrumental resolution function, Lorentzian isotropic size (Y) and Lorentzian isotropic strain (X) parameters were refined but only the second one (strain effect) improved the fits. Then, anisotropic refinements of strain parameters were attempted to get better peak profiles. The corresponding spherical harmonic coefficients are given in the following tables but the estimated standard deviations (esd) are too large to discuss the physical meaning of these values.

The low data to parameter ratio at high pressure - mainly due to the coexistence of starting and HP structure and to the peak broadening increasing with the strain - prevent us from a stable refinement of atomic positions.

Table 1. Results of the Rietveld refinement from X-ray diffraction analysis ($\lambda_{Mo}$) of $ZrS_2$.

| **Pressure 0.37GPa** | | | |
|---|---|---|---|
| **1T-$ZrS_2$** | *P -3m1* | $R_{Bragg}$ = 9.93 % | Fract. (%) = 96 (4) |
| *a* = b = 3.6619 (11) Å | *c* = 5.7746 (24) Å | | |
| *V* = 67.059 (40) Å$^3$ | | | |
| **Strain Parameters (for profile only, large esd)** | | | |
| $S_{400}$=3.21 (2.43), $S_{004}$=11.11 (4.70), $S_{112}$=0 | | | |
| **Fe** | *I m3m* | $R_{Bragg}$ = 11.7 % | Fract. (%) = 4 (1) |
| *a* = 2.8771 (52) Å | | | |
| *V* = 23.816 (74) Å$^3$ | | | |
| **Pressure 2GPa** | | | |
| **1T-$ZrS_2$** | *P -3m1* | $R_{Bragg}$ = 16.3 % | Fract. (%) = 93 (4) |
| *a* = b = 3.6365 (9) Å | *c* = 5.6330 (19) Å | | |
| *V* = 64.511 (31) Å$^3$ | | | |
| **Strain Parameters (for profile only, large esd)** | | | |
| $S_{400}$=1.73 (1.96), $S_{004}$=8.20(+/-3.71), $S_{112}$=0 | | | |
| **Fe** | *I m3m* | $R_{Bragg}$ = 30.2 % | Fract. (%) = 7 (1) |
| *a* = 2.8714 (39) Å | | | |
| *V* = 23.675 (56) Å$^3$ | | | |
| **Pressure 4GPa** | | | |
| **1T-$ZrS_2$** | *P -3m1* | $R_{Bragg}$ = 12.1 % | Fract. (%) = 56 (5) |
| *a* = b = 3.6122 (7) Å | *c* = 5.5275 (26) Å | | |
| *V* = 62.461 (34) Å$^3$ | | | |
| **Strain Parameters (for profile only, large esd)** | | | |
| $S_{400}$=1.25 (0), $S_{004}$=5.67(4.86), $S_{112}$=0.23 (11.30) | | | |
| **HP-$ZrS_2$** | *P $2_1$/m* | $R_{Bragg}$ = 17.8 % | Fract. (%) = 38 (8) |
| *a* = 6.4811 (104) Å | *b* = 3.7610 (10) Å | *c* = 5.0823 (126) Å | β= 108.922(50) deg |
| *V* = 117.188 (341) Å$^3$ | | | |
| **Strain Parameters (for profile only, large esd)** | | | |
| $S_{400}$=45 (83), $S_{040}$=55(66), $S_{004}$=140(306), $S_{220}$=138(127), $S_{202}$=0 | | | |
| $S_{022}$=0, $S_{121}$=-0, $S_{301}$=0, $S_{103}$=0 | | | |
| **Fe** | *I m3m* | $R_{Bragg}$ = 15.0 % | Fract. (%) = 6 (1) |
| *a* = 2.8616 (30) Å | | | |
| *V* = 23.433 (42) Å$^3$ | | | |
| **Pressure 6GPa** | | | |
| **1T-$ZrS_2$** | *P -3m1* | $R_{Bragg}$ = 6.2 % | Fract. (%) = 37 (4) |
| *a* = b = 3.5904 (17) Å | *c* = 5.4493 (52) Å | | |
| *V* = 60.837 (71) $Å^3$ | | | |
| **Strain Parameters (for profile only, large esd)** $S_{400}$=26 (29), $S_{004}$=0 (8), $S_{112}$=0 | | | |
| **HP-$ZrS_2$** | *P $2_1$/m* | $R_{Bragg}$ = 11.6 % | Fract. (%) = 55 (6) |
| *a* = 6.4766 (63) Å | *b* = 3.7235 (25) Å | *c* = 4.9906 (75) Å | β= 108.549(54) deg |
| *V* = 114.100 (218) $Å^3$ | | | |
| **Strain Parameters (for profile only, large esd)** $S_{400}$=45 (28), $S_{040}$=146(52), $S_{004}$=46(60), $S_{220}$=88(90), $S_{202}$=124 (104) $S_{022}$=0, $S_{121}$=-0, $S_{301}$=0, $S_{103}$=0 | | | |
| **Fe** | *I m3m* | $R_{Bragg}$ = 7.1 % | Fract. (%) = 8 (1) |
| *a* = 2.8587 (34) Å | | | |
| *V* = 23.363 (48) $Å^3$ | | | |
| **Pressure 8GPa** | | | |
| **1T-$ZrS_2$** | *P -3m1* | $R_{Bragg}$ = 8.7 % | Fract. (%) = 28 (6) |
| *a* = b = 3.5789 (28) Å | *c* = 5.4162 (99) Å | | |
| *V* = 60.081 (128) $Å^3$ | | | |
| **Strain Parameters (for profile only, large esd)** $S_{400}$=109 (85), $S_{004}$=63 (69), $S_{112}$=126 (138) | | | |
| **HP-$ZrS_2$** | *P $2_1$/m* | $R_{Bragg}$ = 15.0 % | Fract. (%) = 62 (8) |
| *a* = 6.4791 (80) Å | *b* = 3.7064 (41) Å | *c* = 5.0004 (123) Å | β= 108.88(11) deg |
| *V* = 113.620 (337) $Å^3$ | | | |
| **Strain Parameters (for profile only, large esd)** $S_{400}$=64 (26), $S_{040}$=169 (65), $S_{004}$=262 (126), $S_{220}$=162 (109), $S_{202}$=0 $S_{022}$=0, $S_{121}$=-0, $S_{301}$=0, $S_{103}$=0 | | | |
| **Fe** | *I m3m* | $R_{Bragg}$ = 10.1 % | Fract. (%) = 10 (1) |
| *a* = 2.8641 (29) Å | | | |
| *V* = 23.493 (41) $Å^3$ | | | |
| **Pressure 10GPa** | | | |
| **1T-$ZrS_2$** | *P -3m1* | $R_{Bragg}$ = 12.3 % | Fract. (%) = 28 (11) |
| *a* = b = 3.5563 (14) Å | *c* = 5.3731 (90) Å | | |
| *V* = 58.850 (104) $Å^3$ | | | |
| **Strain Parameters (for profile only, large esd)** $S_{400}$=155 (172), $S_{004}$= 383 (299), $S_{112}$= 0 | | | |
| **HP-$ZrS_2$** | *P $2_1$/m* | $R_{Bragg}$ = 15.6 % | Fract. (%) = 65 (9) |
| *a* = 6.4673 (94) Å | *b* = 3.6731 (21) Å | *c* = 4.9983 (138) Å | β= 109.23(13) deg |
| *V* = 112.106 (355) $Å^3$ | | | |
| **Strain Parameters (for profile only, large esd)** $S_{400}$=11 (22), $S_{040}$=200 (64), $S_{004}$=125 (118), $S_{220}$=210 (85), $S_{202}$= 286 (134) $S_{022}$=0, $S_{121}$=-0, $S_{301}$=0, $S_{103}$=0 | | | |
| **Fe** | *I m3m* | $R_{Bragg}$ = 8.9 % | Fract. (%) = 7.1 (2) |
| *a* = 2.8558 (31) Å | | | |
| *V* = 23.290 (44) $Å^3$ | | | |

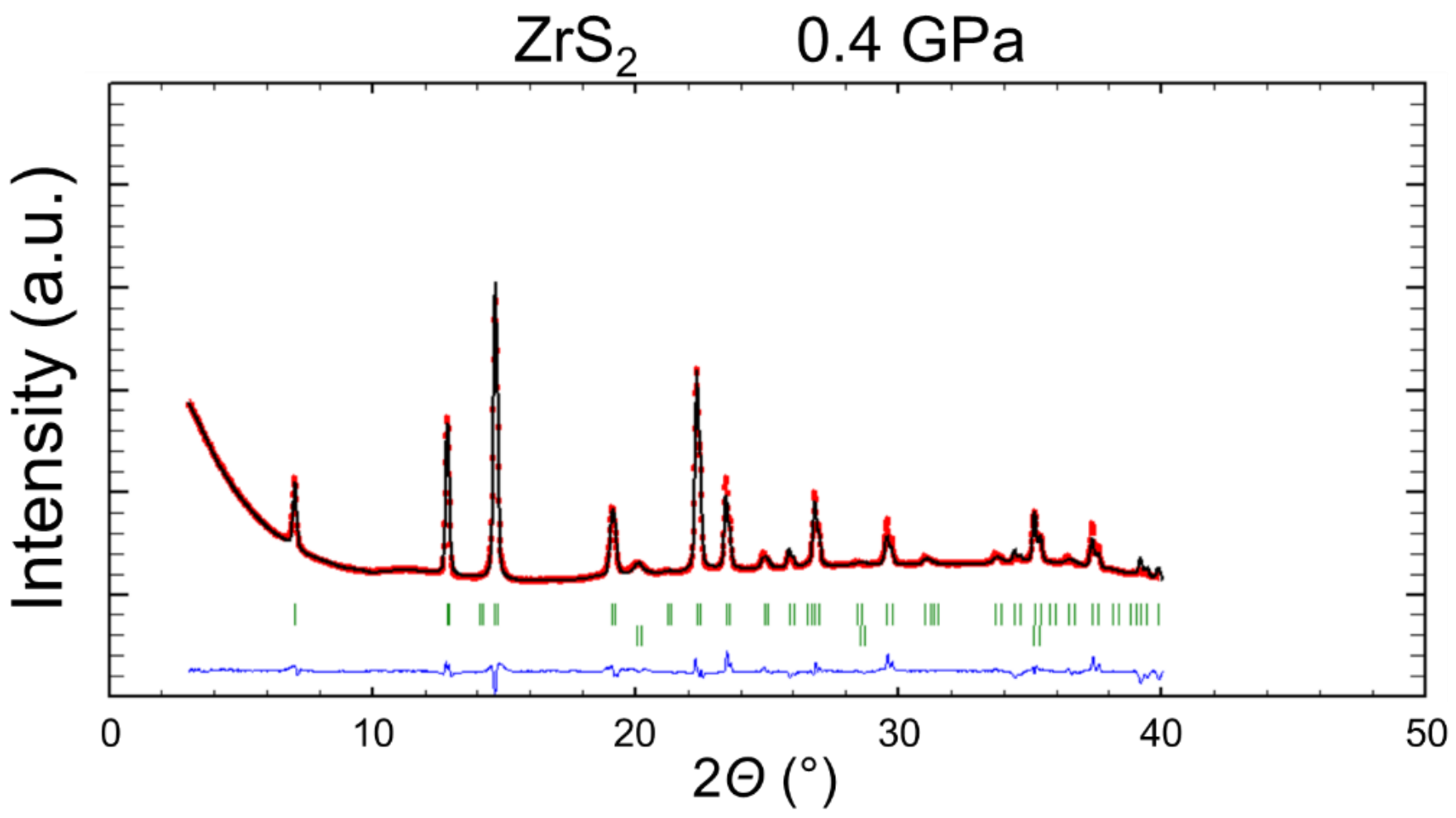

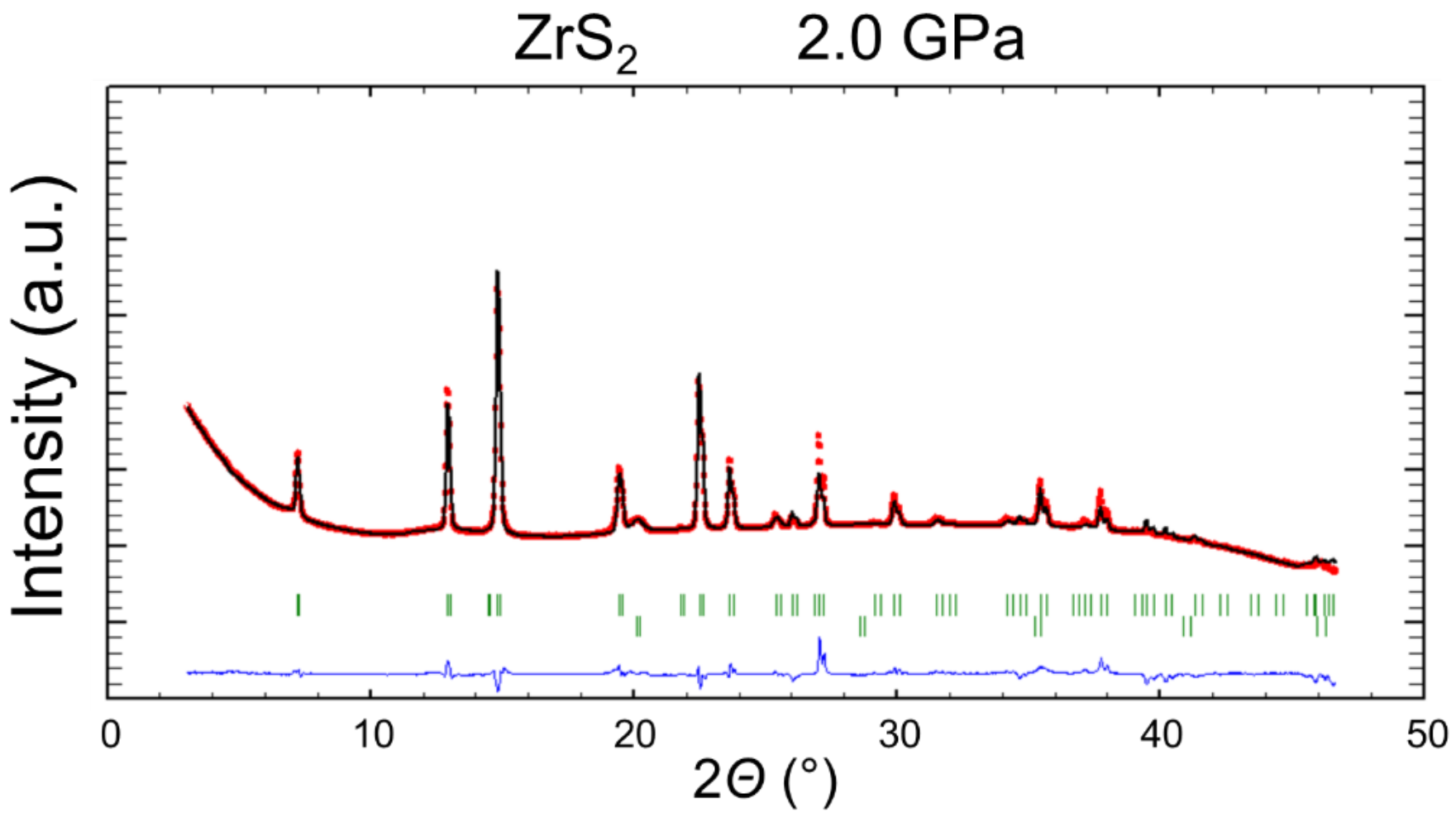

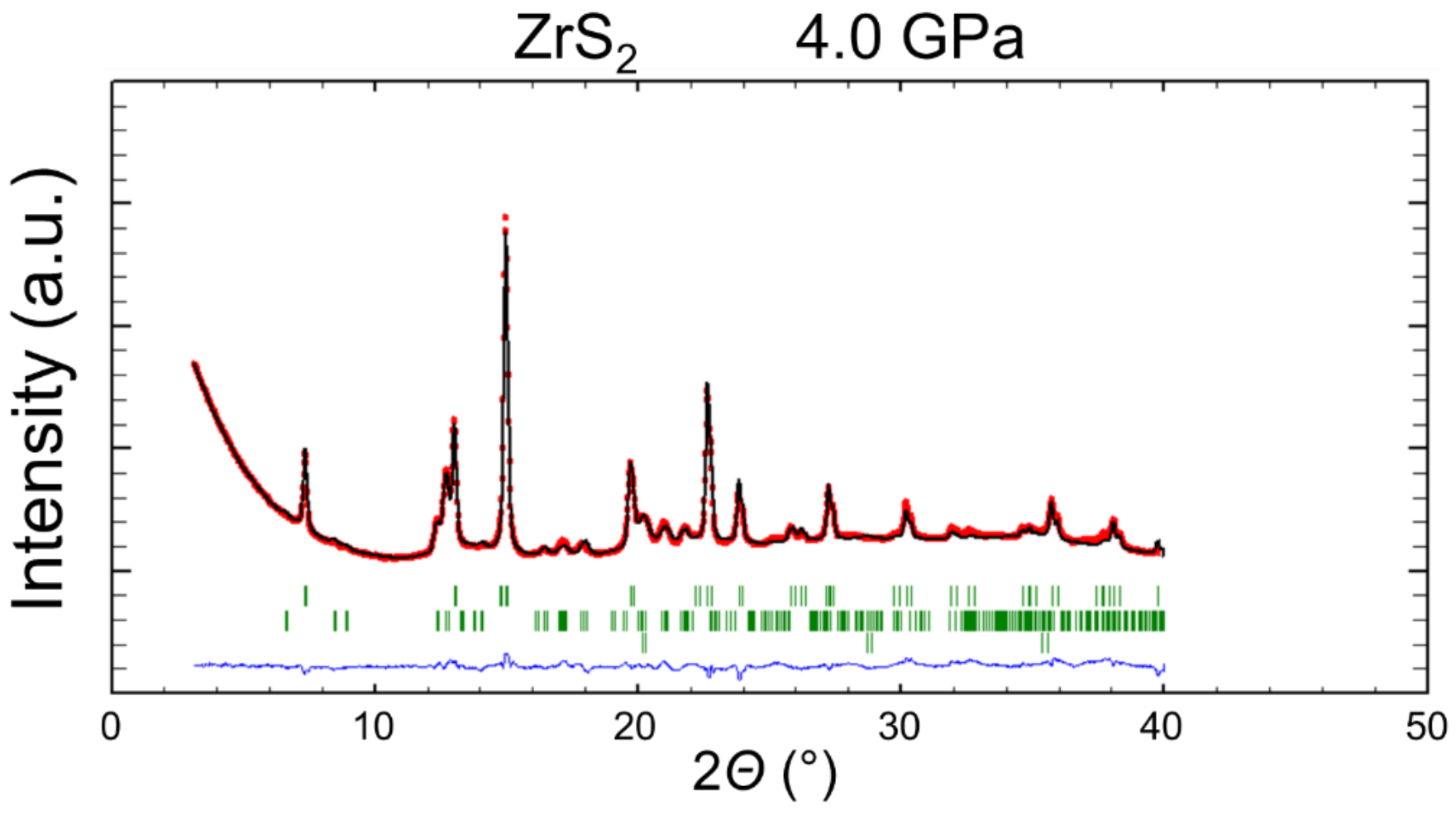

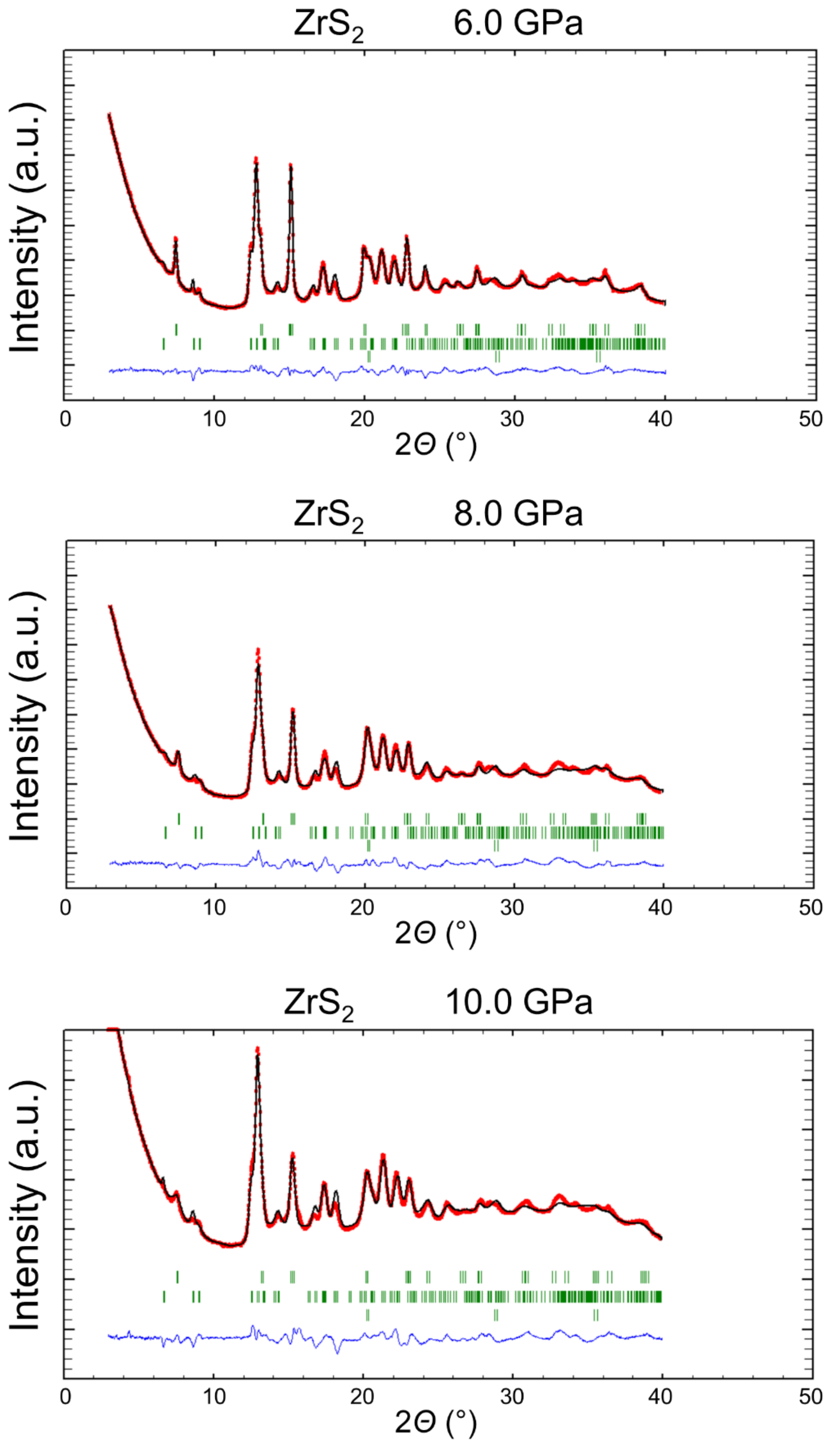

**Figure SI 1** Results of the Rietveld refinement from X-ray diffraction analysis ($\lambda_{Mo}$) of $ZrS_2$.

Table X. Results of the Rietveld refinement from X-ray diffraction analysis ($\lambda_{Mo}$) of $ZrSe_2$.

| | | | |
|---|---|---|---|
| **Pressure 0.4GPa** | | | |
| **1T-ZrSe$_2$** | *P -3m1* | $R_{Bragg}$ = 7.72 % | Fract. (%) = 98 (8) |
| *a* = b = 3.7656 (2) Å | *c* = 6.0953 (30) Å | | |
| *V* = 74.854 (37) Å$^3$ | | | |
| **Strain Parameters (for profile only, large esd)** | | | |
| $S_{400}$=1.16 (34), $S_{004}$=5.76 (4.31), $S_{112}$=3.32 (4.50) | | | |
| **Fe** | *I m3m* | $R_{Bragg}$ = 6.65 % | Fract. (%) = 2 (2) |
| *a* = 2.8756 (8) Å | | | |
| *V* = 23.779 (11) Å$^3$ | | | |
| **Pressure 1.7GPa** | | | |
| **1T-ZrSe$_2$** | *P -3m1* | $R_{Bragg}$ = 8.28 % | Fract. (%) = 98 (10) |
| *a* = b = 3.7335 (13) Å | *c* = 6.0164 (71) Å | | |
| *V* = 72.629 (92) Å$^3$ | | | |
| **Strain Parameters (for profile only, large esd)** | | | |
| $S_{400}$= 0.51 (18), $S_{004}$= 5.68 (4.33), $S_{112}$= 3.19 (4.30) | | | |
| **Fe** | *I m3m* | $R_{Bragg}$ = 7.07% | Fract. (%) = 2 (2) |
| *a* = 2.8486 (10) Å | | | |
| *V* = 23.114 (14) Å$^3$ | | | |
| **Pressure 3.3GPa** | | | |
| **1T-ZrSe$_2$** | *P -3m1* | $R_{Bragg}$ = 10.5 % | Fract. (%) = 99 (7) |
| *a* = b = 3.7213 (2) Å | *c* = 5.8494 (24) Å | | |
| *V* = 70.150 (29) Å$^3$ | | | |
| **Strain Parameters (for profile only, large esd)** | | | |
| $S_{400}$= 1.20 (27), $S_{004}$= 4.11 (3.13), $S_{112}$= 0.22 (3.07) | | | |
| **Fe** | *I m3m* | $R_{Bragg}$ = 4.07 % | Fract. (%) = 1 (3) |
| *a* = 2.8215 (56) Å | | | |
| *V* = 22.462 (8) Å$^3$ | | | |
| **Pressure 5.2GPa** | | | |
| **1T-ZrSe$_2$** | *P -3m1* | $R_{Bragg}$ = 14.4 % | Fract. (%) = 99 (6) |
| *a* = b = 3.6909 (2) Å | *c* = 5.7716 (27) Å | | |
| *V* = 68.090 (33) Å$^3$ | | | |
| **Strain Parameters (for profile only, large esd)** | | | |
| $S_{400}$= 15.32 (2.45), $S_{004}$= 55.78 (14.19), $S_{112}$=0 | | | |
| **Fe** | *I m3m* | $R_{Bragg}$ = 3.90 % | Fract. (%) = 1(1) |
| *a* = 2.7942 (9) Å | | | |
| *V* = 21.817 (12) Å$^3$ | | | |
| **Pressure 7.4GPa** | | | |
| **1T-$ZrSe_2$** | *P -3m1* | $R_{Bragg}$ = 10.2 % | Fract. (%) = 58 (5) |
| *a* = b = 3.6473 (9) Å | *c* = 5.7208 (1) Å | | |
| *V* = 65.907 (24) $Å^3$ | | | |
| **Strain Parameters (for profile only, large esd)** | | | |
| NR | | | |
| **HP-$ZrSe_2$** | *I mmm* | $R_{Bragg}$ = 14.3 % | Fract. (%) = 41 (2) |
| *a* = 3.6097 (20) Å | *b* = 5.3843 (82) Å | *c* = 12.1040 (103) Å | |
| *V* = 235.252 (432) $Å^3$ | | | |
| **Strain Parameters (for profile only, large esd)** | | | |
| $S_{400}$= 6.98(7.71), $S_{040}$= 21.10 (31.36), $S_{004}$= 1.43 (1.44), $S_{220}$= 10.68 (72.44), $S_{202}$=0, $S_{022}$=0 | | | |
| **Fe** | *I m3m* | $R_{Bragg}$ = 3.16 % | Fract. (%) = 1(1) |
| *a* = 2.7664 (0.0032) Å | | | |
| *V* = 21.171 (42) $Å^3$ | | | |
| **Pressure 9.5GPa** | | | |
| **1T-$ZrSe_2$** | *P -3m1* | $R_{Bragg}$ = 13.3 % | Fract. (%) = 39 (5) |
| *a* = b = 3.6189 (16) Å | *c* = 5.7208 (2) Å | | |
| *V* = 64.886 (41) $Å^3$ | | | |
| **Strain Parameters** | | | |
| NR | | | |
| **HP-$ZrSe_2$** | *I mmm* | $R_{Bragg}$ = 13.9 % | Fract. (%) = 61 (8) |
| *a* = 3.5824 (21) Å | *b* = 5.3699 (84) Å | *c* = 12.0613 (233) Å | |
| *V* = 232.023 (592) $Å^3$ | | | |
| **Strain Parameters (for profile only, large esd)** | | | |
| $S_{400}$= 0.48 (41), $S_{040}$= 1.03 (1.55), $S_{004}$= 0.09 (7), $S_{220}$= 1.55 (3.81), $S_{202}$= 0, $S_{022}$=0 | | | |
| **Fe** | *I m3m* | $R_{Bragg}$ = 0.5 % | Fract. (%) = 0 (1) |
| *a* = 2.7413 (497) Å | | | |
| *V* = 20.599 (647) $Å^3$ | | | |
| **Pressure 11.5GPa** | | | |
| **1T-$ZrS_2$** | *P -3m1* | $R_{Bragg}$ = 17.1 % | Fract. (%) = 19(3) |
| *a* = b = 3.6307 (22) Å | *c* = 5.6089 (501) Å | | |
| *V* = 64.031 (574) $Å^3$ | | | |
| **Strain Parameters** | | | |
| NR | | | |
| **HP-$ZrS_2$** | *I mmm* | $R_{Bragg}$ = 15.4 % | Fract. (%) = 81 (10) |
| *a* = 3.5894 (239) Å | *b* = 5.4119 (111) Å | *c* = 12.0219 (116) Å | |
| *V* = 233.527 (552) $Å^3$ | | | |
| **Strain Parameters (for profile only, large esd)** | | | |
| $S_{400}$= 1.09(0.74), $S_{040}$= 2.03 (2.72), $S_{004}$= 0.07 (0.08), $S_{220}$= NR, $S_{202}$= 0.67 (0.68), $S_{022}$=0 | | | |
| **Fe**<br>**not refined**<br>**(too much overlap)** | *I m3m* | $R_{Bragg}$ = NR | Fract. (%) = NR |
| *a* = NR | | | |
| *V* = NR | | | |

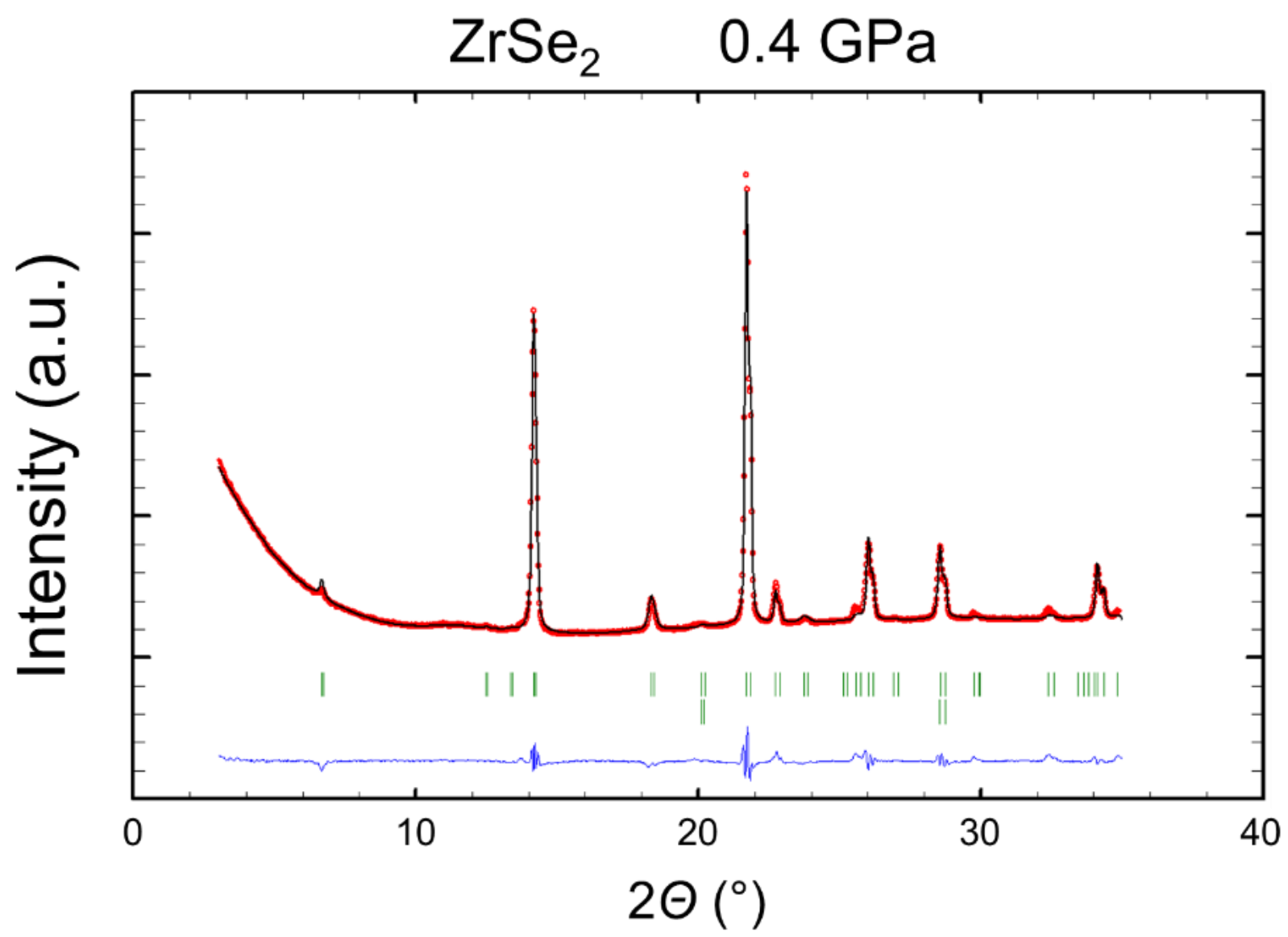

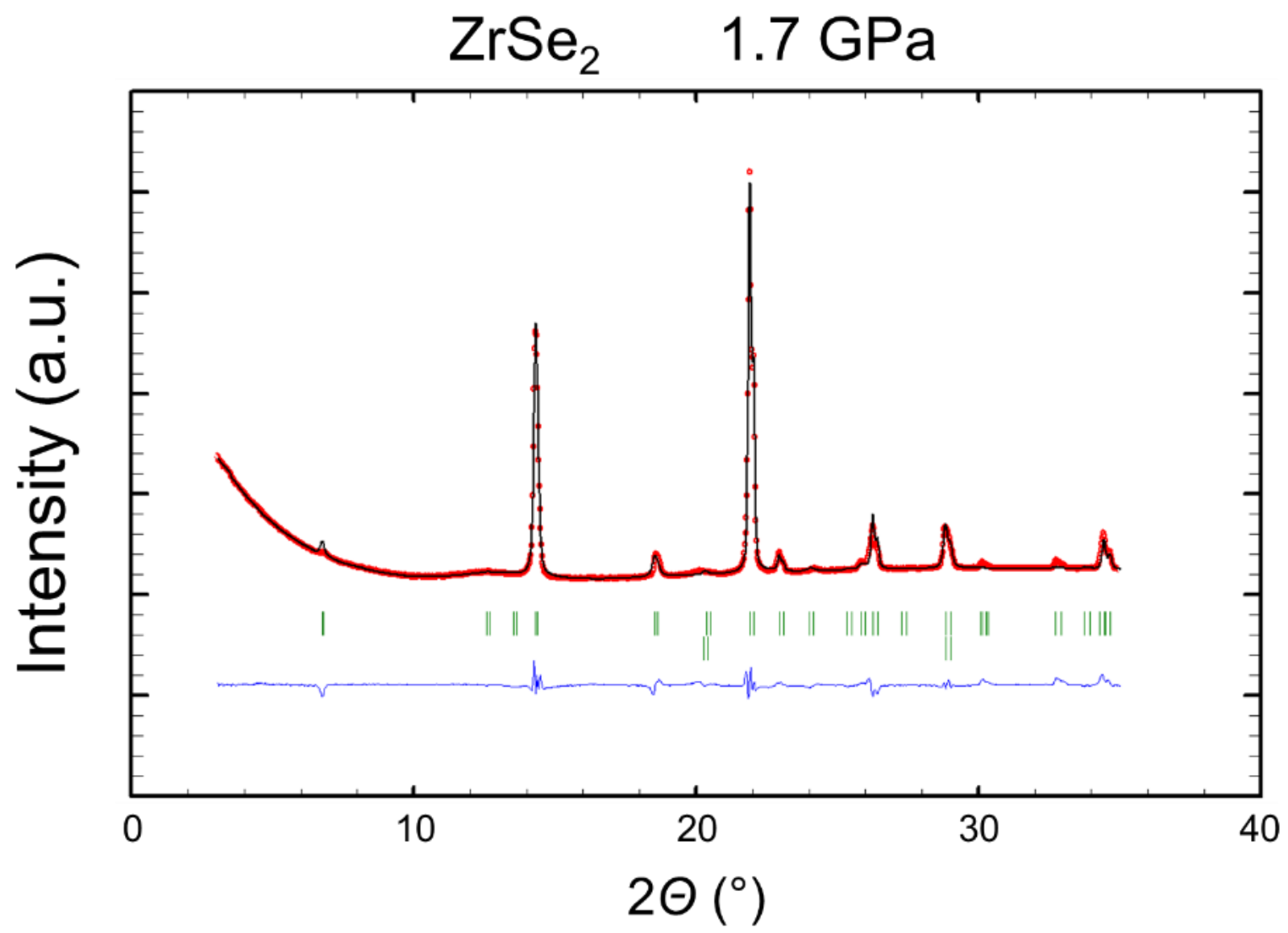

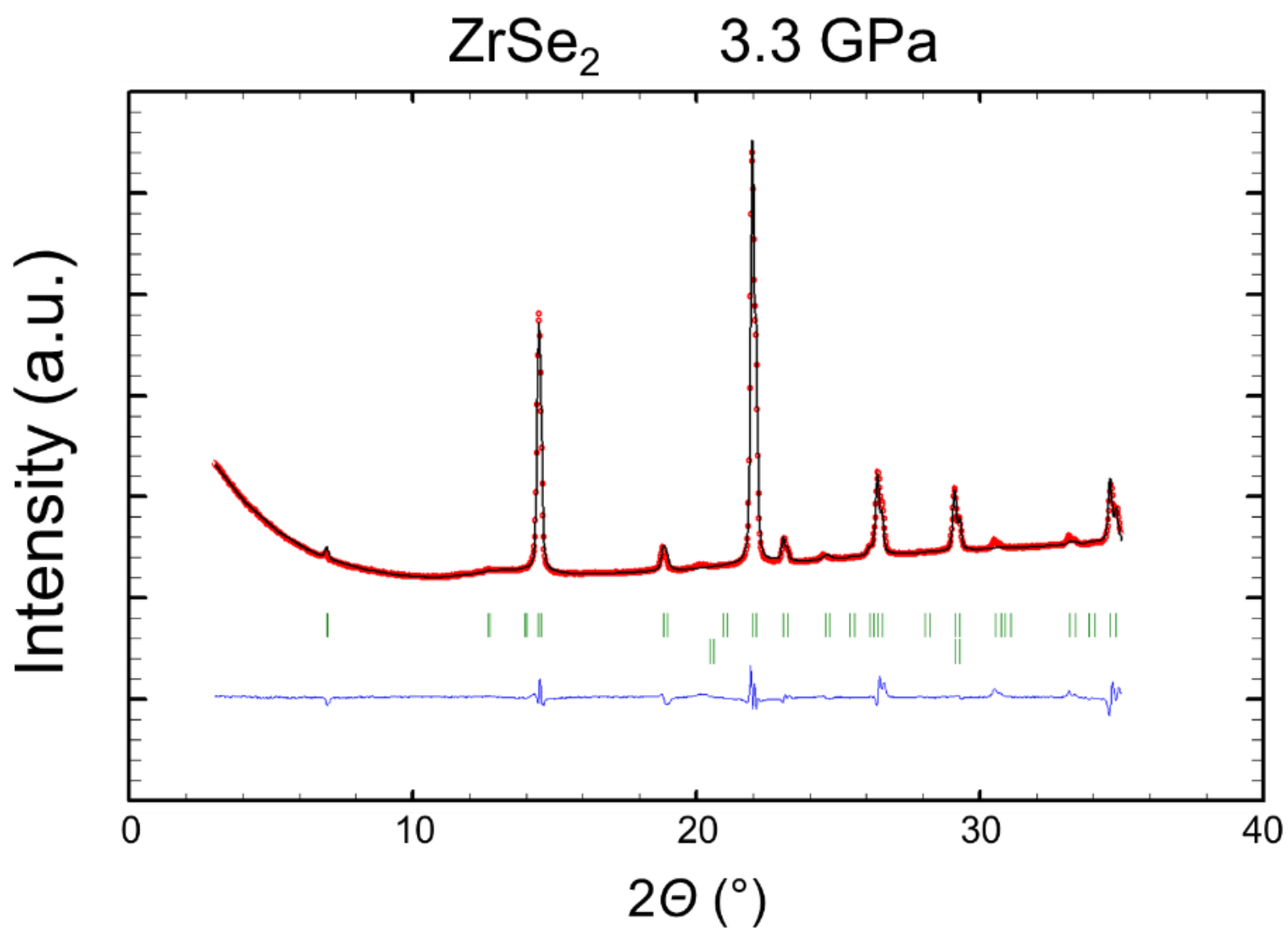

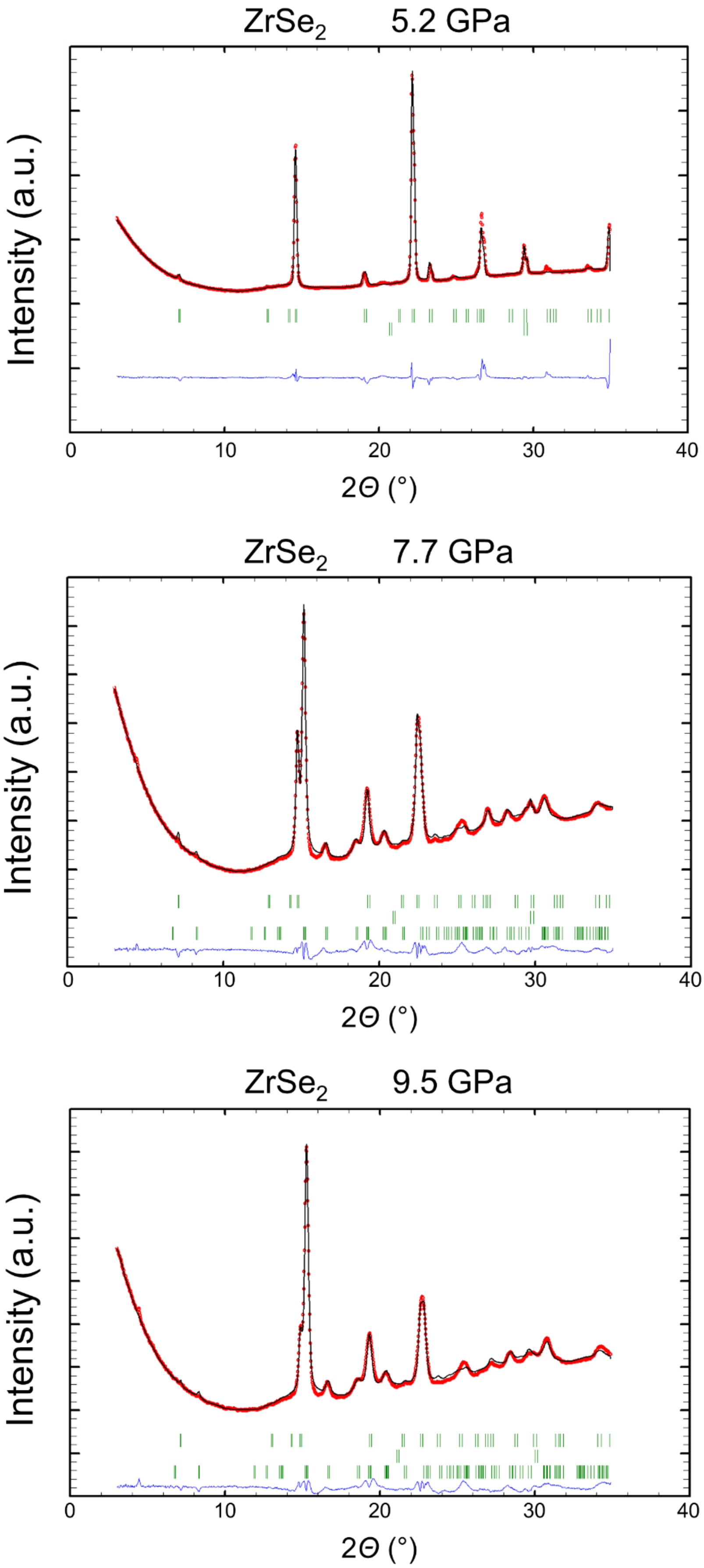

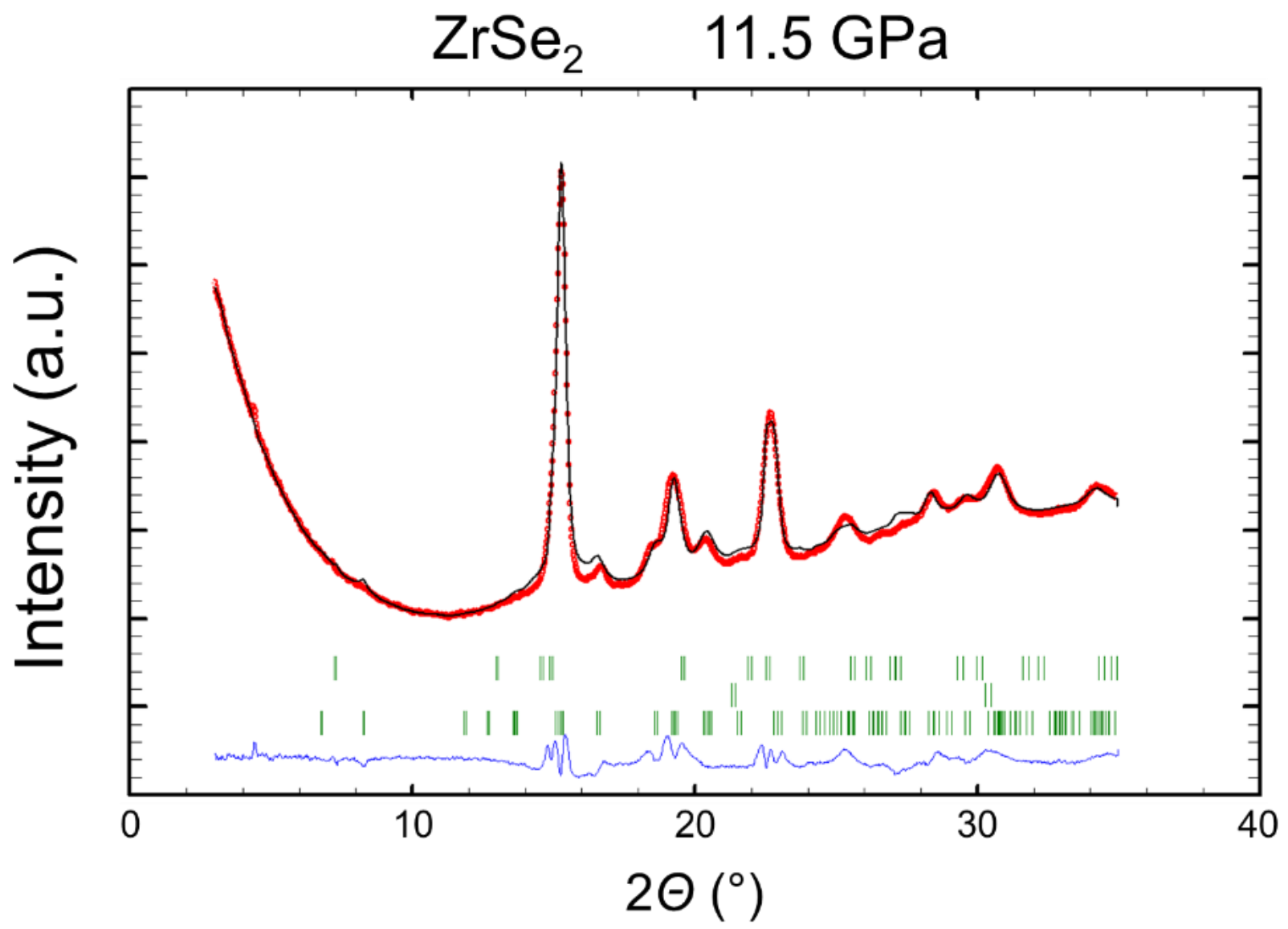

**Figure SI 2** Results of the Rietveld refinement from X-ray diffraction analysis ($\lambda_{Mo}$) of $ZrS_2$.

## Pressure - Strain relation

Given the experimental X-ray diffraction (XRD) data, recorded during high-pressure experiments, it is possible to compute the change in unit cell parameters of the 1T-phases of $ZrS_2$ and $ZrSe_2$. The data were used to compute the pressure – strain relation for the material. The engineering strain (in percentage) applied to the material is computed as:

$$\varepsilon_a(P) = \frac{a(P)-a(0)}{a(0)} \cdot 100 \qquad (1)$$

Where $a(0)$ and $a$(P) are the same lattice parameter at ambient and pressure P respectively. Given the hydrostatic conditions of the experiment, only the three diagonal component of the strain tensor are non-zero: $\varepsilon_a$, $\varepsilon_b$ and $\varepsilon_c$. Because of the hexagonal lattice symmetry (P-3m1),

$$\varepsilon_a=\varepsilon_b. \qquad (2)$$

As a reference, strain is considered positive when tensile and negative when compressive. In the case of our high-pressure experiments, ε is negative as only compressive deformation is produced. In Figure SI 1, we report the computed values, extracted from high-pressure powder XRD for 1T-$ZrS_2$ and 1T-$ZrSe_2$. As a comparison, Figure SI 2 shows the compressibility as function of pressure for the semiconducting 2H polytypes ($MoS_2$, $WS_2$ and $WSe_2$), computed from published results [2,3,4].

From our results the 1T polytypes appear to have a much higher compressibility than the 2H polytypes. In comparison, at 10 GPa, in-plane strain are $\varepsilon_a$=3.6% and $\varepsilon_a$=2.2% for 1T and 2H respectively. For both materials, the out-of-plane compressibility is much higher than the in-plane one, as one would expect considering the much weaker van der Waals interaction.

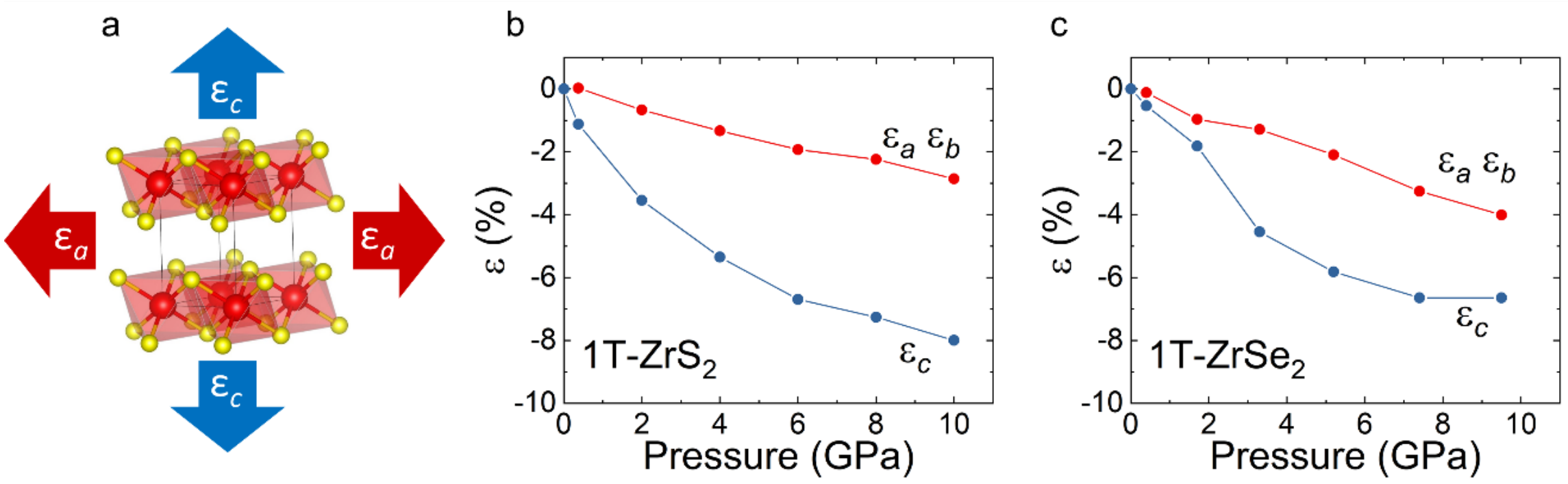

**Figure SI 3.** **a**) 1T-$ZrX_2$ unit cell. The arrows point in the direction of positive strains (tensile) along 2 principal directions, $\varepsilon_a$ is in-plane and $\varepsilon_c$ is out-of-plane. **b**) Experimentally determined pressure-strain relation for 1T-$ZrS_2$. **c**) Experimentally determined pressure-strain relation for 1T-$ZrSe_2$.

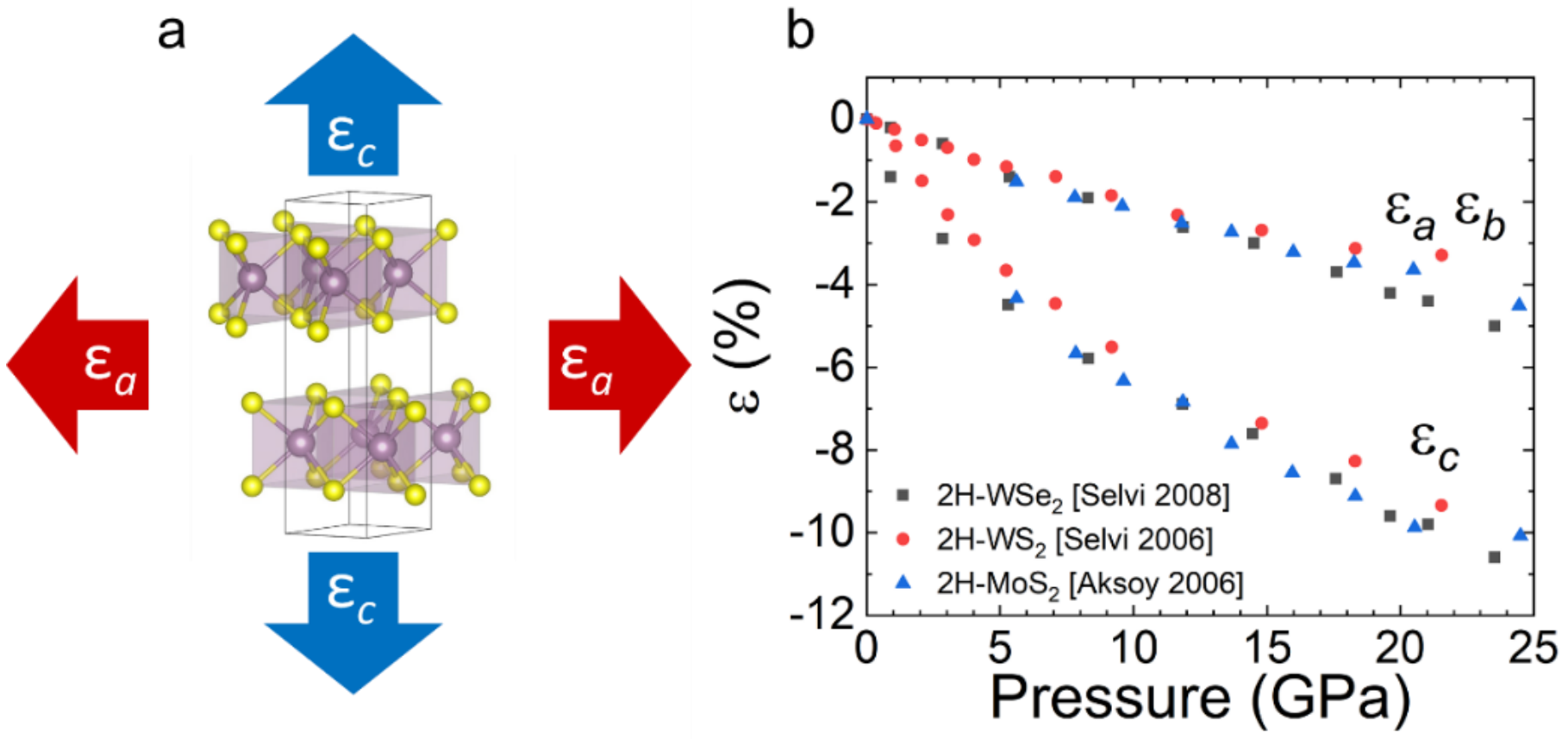

**Figure SI 4.** **a**) 2H-$MeX_2$ unit cell. The arrows point in the direction of positive strains (tensile) along 2 principal directions. **b**) Experimentally determined pressure-strain relation for 2H-$MoS_2$, 2H-$WS_2$ and 2H-$WSe_2$.

## High-pressure Raman scattering

Here we show the Raman spectra at each pressure during compression and decompression.

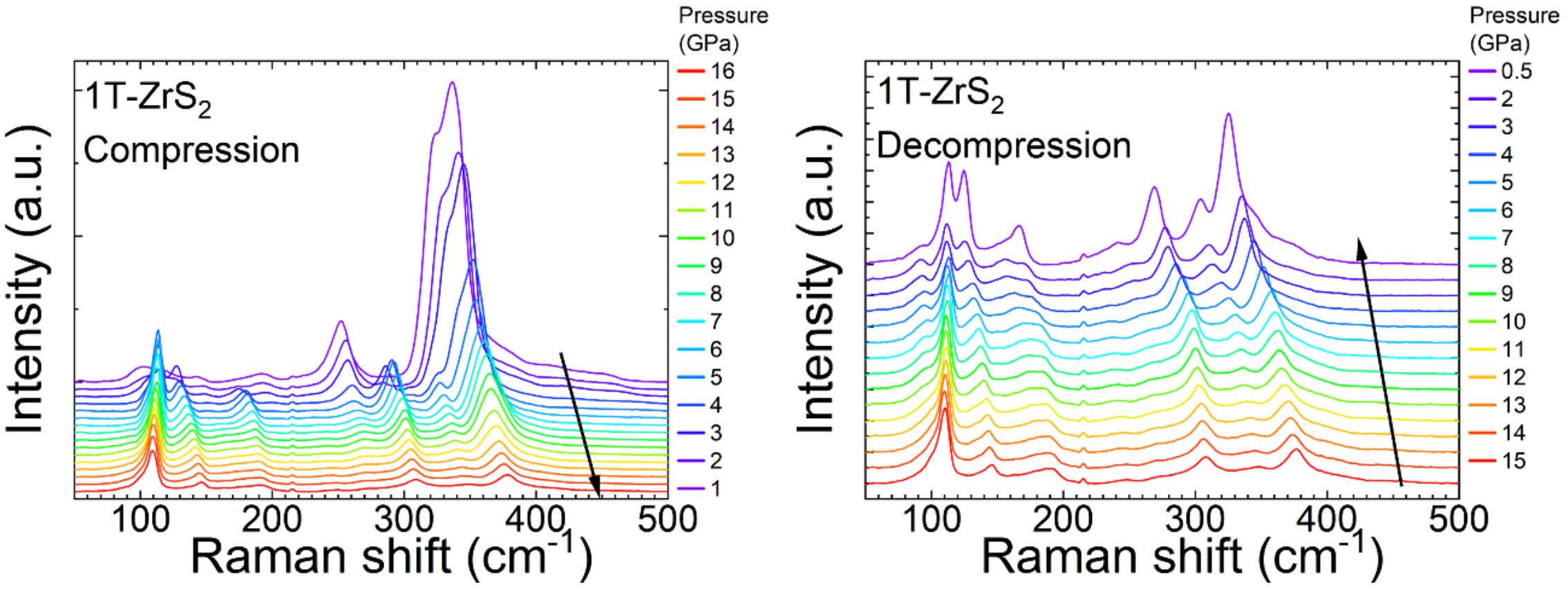

**Figure SI 5.** Raman spectra of $ZrS_2$ collected during compression and decompression. All curves are shifted vertically by an equal amount.

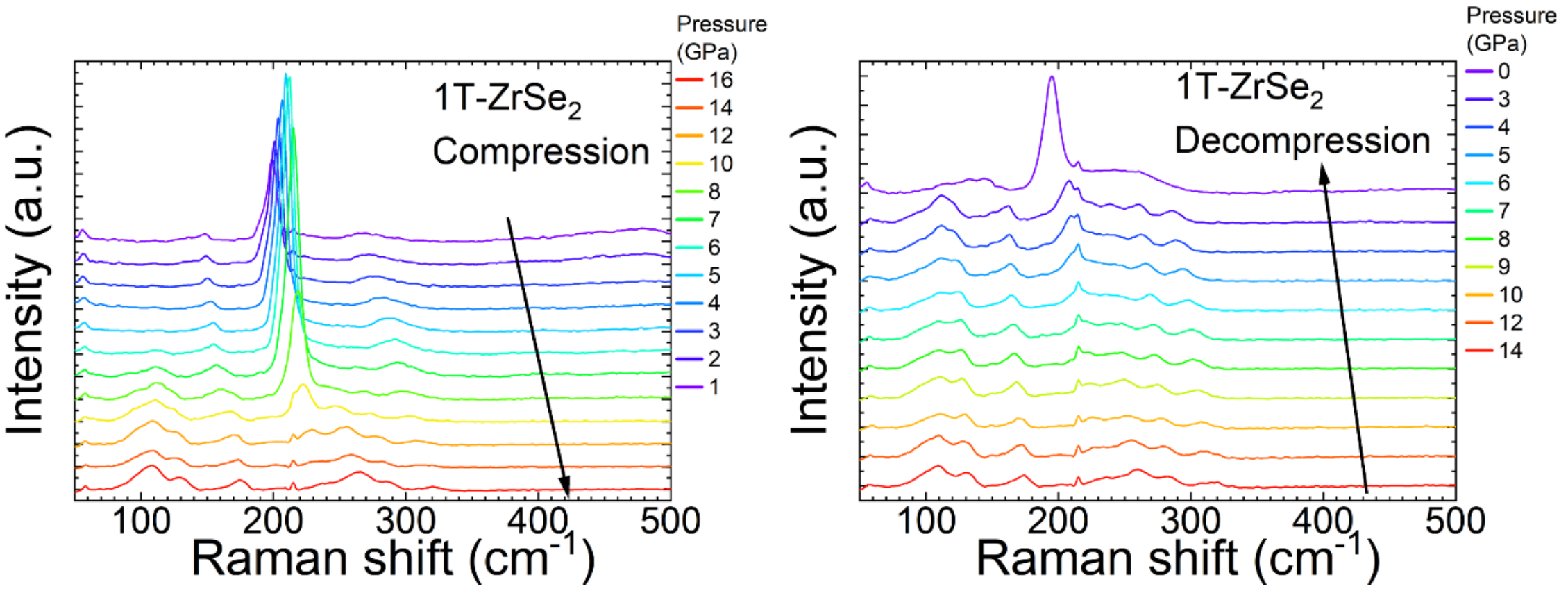

**Figure SI 6.** Raman spectra of $ZrSe_2$ collected during compression and decompression. All curves are shifted vertically by an equal amount.

## Band gap estimation

Values of the optical band gap as a function of pressure were estimated from the optical transmittance data. As all semiconducting phases have an indirect band gap, we used the same empirical model for all data points. After computing the absorption coefficient (α) from the experimental data, we estimated the indirect band gap $E_g$ via linear extrapolations of $(h\vartheta\alpha)^{1/2}$. Representative fitting are shown in Figure SI. 7.

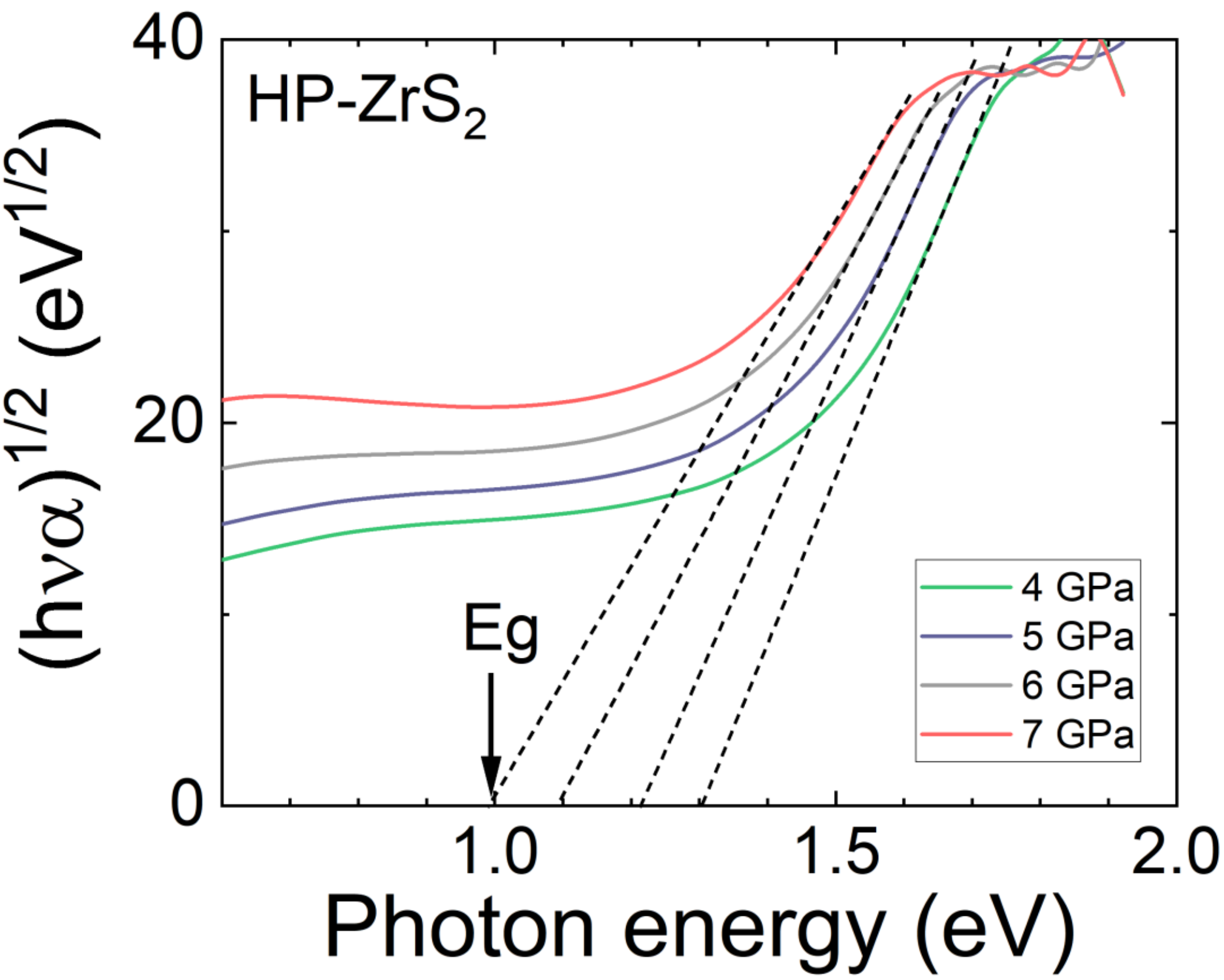

**Figure SI 7.** Representative extrapolation of the indirect band gap for the high-pressure phase of $ZrS_2$ (HP-$ZrS_2$) at different pressures.

## DFT band structures calculations

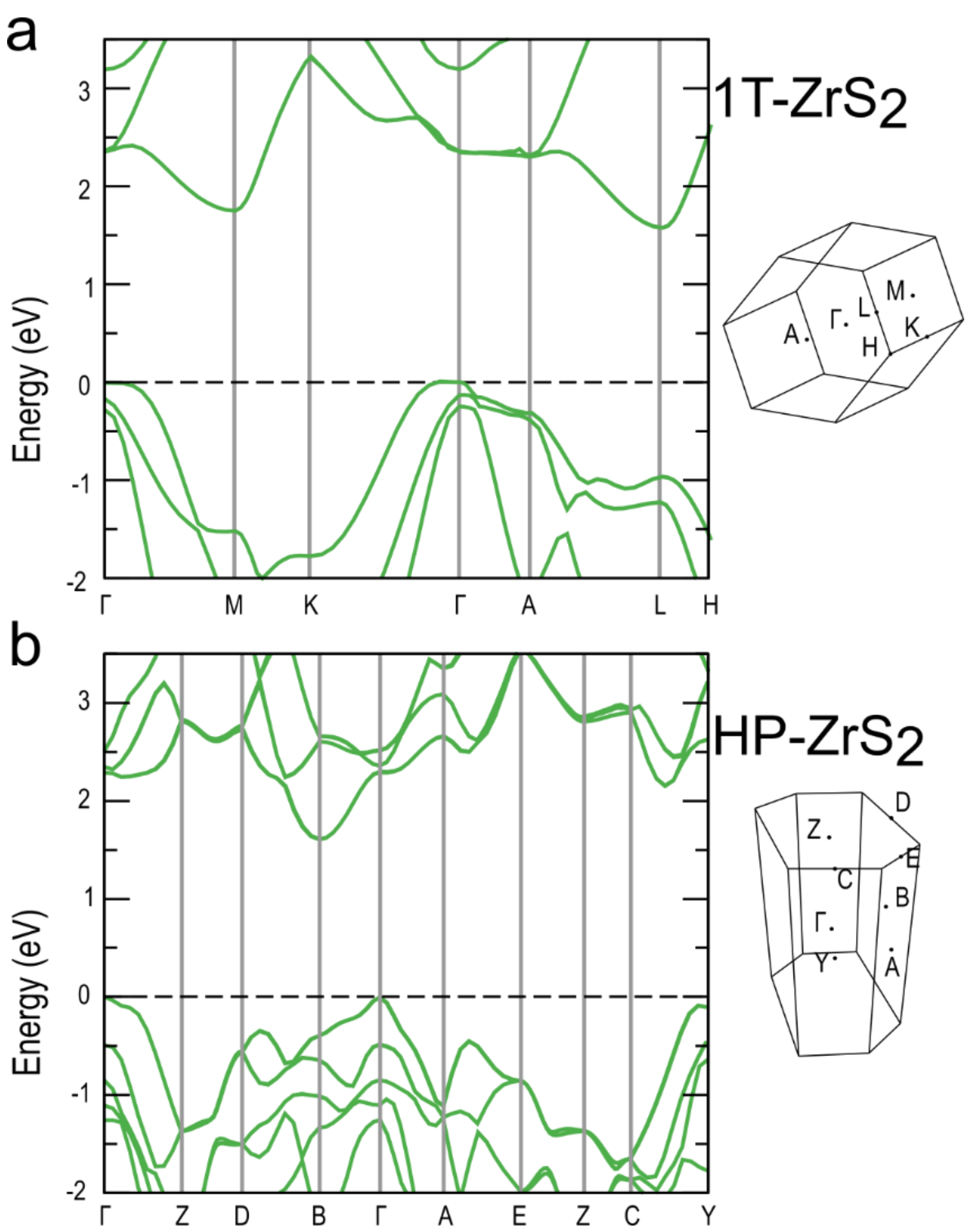

**Figure SI 8.** Electronic band structure of the (a) low-pressure (0 GPa) and (b) high-pressure (6 GPa) crystalline phases of $ZrS_2$, together with the high-symmetry points in the corresponding first Brillouin zones.

## References

[1] Rodríguez-Carvajal, Juan. "Recent advances in magnetic structure determination by neutron powder diffraction." *Physica B*. **1993**, 192.1-2: 55-69.

[2] Aksoy, R. et al. X-ray diffraction study of molybdenum disulfide to 38.8 GPa. *J. Phys. Chem. Solids*. **2006**, 67, 1914–1917.

[3] Selvi, E., Ma, Y., Aksoy, R., Ertas, A. & White, A. High pressure X-ray diffraction study of tungsten disulfide. *J. Phys. Chem. Solids*. **2006**, 67, 2183–2186.

[4] Selvi, E., Aksoy, R., Knudson, R. & Ma, Y. High-pressure X-ray diffraction study of tungsten diselenide. *J. Phys. Chem. Solids*. **2008**, 69, 2311–2314.